\documentclass[twoside,twocolumn,9pt]{article}
\usepackage{extsizes}
\usepackage[super,sort&compress,comma]{natbib} 
\usepackage[version=3]{mhchem}
\usepackage[left=1.5cm, right=1.5cm, top=1.785cm, bottom=2.0cm]{geometry}
\usepackage{balance}
\usepackage{mathptmx}
\usepackage{sectsty}
\usepackage{graphicx} 
\usepackage{lastpage}
\usepackage[format=plain,justification=justified,singlelinecheck=false,font={stretch=1.125,small,sf},labelfont=bf,labelsep=space]{caption}
\usepackage{float}
\usepackage{fancyhdr}
\usepackage{fnpos}
\usepackage[english]{babel}
\addto{\captionsenglish}{%
  
}
\usepackage{array}
\usepackage{droidsans}
\usepackage{charter}
\usepackage[T1]{fontenc}
\usepackage[usenames,dvipsnames]{xcolor}
\usepackage{setspace}
\usepackage[compact]{titlesec}
\usepackage{hyperref}
\usepackage{physics}
\usepackage{subcaption}
\usepackage{epstopdf}
\usepackage{amsmath,amssymb}
\usepackage{color}
\usepackage{dcolumn}
\usepackage{bm}
\usepackage{comment}
\definecolor{cream}{RGB}{222,217,201}

    \newcommand{\lsi}{Laboratoire des Solides Irradi\'es, \'Ecole Polytechnique, CNRS, CEA/DRF/IRAMIS, Institut Polytechnique de Paris, F-91128 Palaiseau, France.}
	\newcommand{\etsf}{European Theoretical Spectroscopy Facility (ETSF).}
	\newcommand{\soleil}{Synchrotron SOLEIL, L'Orme des Merisiers, Saint-Aubin, BP 48, F-91192 Gif-sur-Yvette, France.}

\begin{document}

\pagestyle{fancy}
\thispagestyle{plain}
\fancypagestyle{plain}{
\renewcommand{\headrulewidth}{0pt}
}

\makeFNbottom
\makeatletter
\renewcommand\LARGE{\@setfontsize\LARGE{15pt}{17}}
\renewcommand\Large{\@setfontsize\Large{12pt}{14}}
\renewcommand\large{\@setfontsize\large{10pt}{12}}
\renewcommand\footnotesize{\@setfontsize\footnotesize{7pt}{10}}
\makeatother

\renewcommand{\thefootnote}{\fnsymbol{footnote}}
\renewcommand\footnoterule{\vspace*{1pt}%
\color{cream}\hrule width 3.5in height 0.4pt \color{black}\vspace*{5pt}} 
\setcounter{secnumdepth}{5}

\makeatletter 
\renewcommand\@biblabel[1]{#1}            
\renewcommand\@makefntext[1]%
{\noindent\makebox[0pt][r]{\@thefnmark\,}#1}
\makeatother 
\renewcommand{\figurename}{\small{Fig.}~}
\sectionfont{\sffamily\Large}
\subsectionfont{\normalsize}
\subsubsectionfont{\bf}
\setstretch{1.125} 
\setlength{\skip\footins}{0.8cm}
\setlength{\footnotesep}{0.25cm}
\setlength{\jot}{10pt}
\titlespacing*{\section}{0pt}{4pt}{4pt}
\titlespacing*{\subsection}{0pt}{15pt}{1pt}

\fancyfoot{}

\fancyfoot[RO]{\footnotesize{\sffamily{1--\pageref{LastPage} ~\textbar  \hspace{2pt}\thepage}}}
\fancyfoot[LE]{\footnotesize{\sffamily{\thepage~\textbar\hspace{3.45cm} 1--\pageref{LastPage}}}}
\fancyhead{}
\renewcommand{\headrulewidth}{0pt} 
\renewcommand{\footrulewidth}{0pt}
\setlength{\arrayrulewidth}{1pt}
\setlength{\columnsep}{6.5mm}
\setlength\bibsep{1pt}

\makeatletter 
\newlength{\figrulesep} 
\setlength{\figrulesep}{0.5\textfloatsep} 

\newcommand{\topfigrule}{\vspace*{-1pt}%
\noindent{\color{cream}\rule[-\figrulesep]{\columnwidth}{1.5pt}} }

\newcommand{\botfigrule}{\vspace*{-2pt}%
\noindent{\color{cream}\rule[\figrulesep]{\columnwidth}{1.5pt}} }

\newcommand{\dblfigrule}{\vspace*{-1pt}%
\noindent{\color{cream}\rule[-\figrulesep]{\textwidth}{1.5pt}} }

\newcommand{\refeq}[1]{Eq.~\ref{#1}}
\newcommand{\refEq}[1]{Eq.~\ref{#1}}
\newcommand{\refeqs}[1]{Eqs.~\ref{#1}}
\newcommand{\refsec}[1]{Sec.~\ref{#1}}
\newcommand{\qP} {quasi-particle}
\newcommand{\ie}{{i.e.,\ }}
\newcommand{\eg}{{e.g.,\ }}
\newcommand{\nrho}{n}
\newcommand{\reffig}[1]{Fig.~\ref{#1}}
\newcommand{\refFig}[1]{Fig.~\ref{#1}}
\newcommand{\GF}{Green's function}
\newcommand{\ham}{hamiltonian}
\newcommand{\rij}{\mbox{${r_{ij}}$}}
\newcommand{\modrij}{\mbox{${|{\bf r}_{i} - {\bf r}_{j}|}$}}
\newcommand{\modrijt}{\mbox{${|\tilde{{\bf r}}_{i} - \tilde{{\bf r}}_{j}|}$}}
\newcommand{\modrrp}{\mbox{${|{\bf r} - {\bf r}^{\prime}|}$}}
\newcommand{\modriI}{\mbox{${|{\bf r}_{i} - {\bf R}_{I}|}$}}
\newcommand{\modRIJ}{\mbox{${|{\bf R}_{I} - {\bf R}_{J}|}$}}

\newcommand{\rmb}{\mbox{$\{ \vecr_{i} \} $}}
\newcommand{\Rmb}{\mbox{$\{ \vecR_{I} \} $}}

\newcommand{\rp}{\mbox{${r^{\prime}}$}}
\newcommand{\vecrp}{\mbox{${{\bf r}^{\prime}}$}}
\newcommand{\qp}{\mbox{${q^{\prime}}$}}
\newcommand{\vecqp}{\mbox{${{\bf q}^{\prime}}$}}
\newcommand{\kp}{{k^{\prime}}}
\newcommand{\veckp}{\mbox{${{\bf k}^{\prime}}$}}
\newcommand{\bDelta}{\mbox{$\bf \Delta$}}
\newcommand{\bG}{{\bf G}}
\newcommand{\bGp}{\mbox{$\bG^{\prime}$}}
\newcommand{\bk}{{\bf k}}
\newcommand{\bnabla}{\mbox{\boldmath $\nabla$}}
\newcommand{\bq}{{\bf q}}
\newcommand{\bR}{{\bf R}}
\newcommand{\br}{{\bf r}}
\newcommand{\brp}{\mbox{$\br^{\prime}$}}
\newcommand{\brpp}{\mbox{$\br^{\prime\prime}$}}
\newcommand{\brppp}{\mbox{$\br^{\prime\prime\prime}$}}
\newcommand{\bx}{{\bf x}}
\newcommand{\half}{{1 \over 2}}
\newcommand{\bfr}{{\bf r}}
\newcommand{\bfR}{{\bf R}}
\newcommand{\BO}{Born--Oppenheimer}
\newcommand{\xc}{exchange-correlation}
\newcommand{\sgn}{{\rm sgn}}
 \newcommand{\VI}{\cite{martin04x}}
 \newcommand{\mul}{\times}
 \newcommand{\mo}{^{\rm m}}
 \newcommand{\ex}{_{\rm ext}}
\newcommand{\luc}[1]{\textcolor{red}{#1}}
\def\s{\Sigma}
\def\sxc{\Sigma_{\rm xc}}
\def\lir{\bm\tilde L}

\newcommand{\mat}[1]{\textcolor{purple}{#1}}

\makeatother

\twocolumn[
  \begin{@twocolumnfalse}
\par
\vspace{1em}
\sffamily
\begin{tabular}{m{.5cm} p{17cm} }

& \noindent\Huge{\textbf{Strategies to build functionals of the density, or functionals of Green's functions: what can we learn?}} \\
\vspace{0.3cm} & \vspace{0.3cm} \\

 & \noindent\large{Ayoub Aouina\textit{$^{a,b}$}, 
                   Matteo Gatti\textit{$^{a,b,c}$}, and
                   Lucia Reining\textit{$^{a,b}$} 
} \\
\vspace{0.3cm} & \vspace{0.3cm} \\
& \noindent\normalsize{The many-body problem can in general not be solved exactly, and one of the most prominent approximations is to build perturbation expansions. A huge variety of expansions is possible, which differ by the quantity to be expanded, the expansion variable, the starting point, and ideas how to resum or terminate the series. Although much has been discussed and much has been done, some choices were made for historical reasons, in particular, limited computation or storage capacities. The present work aims at examining the justifications for different choices made in different contexts, by comparing ingredients of functionals based on \GF s on one side, and on the charge density on the other side. Of particular interest will be the question of how to build an optimal starting point for the approximation of non-local quantities, making use of near- or far-sightedness, and daring to consider models beyond the homogeneous electron gas. This will include the use of connector approximations. We will also discuss why it is a good idea to build functionals of the density.  
}

\end{tabular}

 \end{@twocolumnfalse} \vspace{0.6cm}

  ]

\renewcommand*\rmdefault{bch}\normalfont\upshape
\rmfamily
\section*{}
\vspace{-1cm}


\footnotetext{\textit{$^{a}$~\lsi}}
\footnotetext{\textit{$^{b}$~\etsf}}
\footnotetext{\textit{$^{c}$~\soleil}}




\section{Introduction}

The general aim of solving the many-body Schr\"odinger equation is to calculate an observable $O$. One could, in principle, obtain $O$ as expectation value involving many-body wavefunctions, but in order to describe realistic materials, one has to find a simpler route. One prominent way to go is to express the expectation value as functional of a quantity $Q$ that is simpler than the many-body wavefunction, $O[Q]$. This raises two problems: first, it is generally not easy to find the functional $O[Q]$, or good approximations to it, and second, $Q$ itself may be an expectation value that is not known a priori. Only if $Q$ is the external potential do we not have to bother about the second question. Otherwise, in the frameworks used here one can obtain $Q$ from an auxiliary system, with an auxiliary potential $v_{\rm aux}$. Again, however, this auxiliary potential itself is unknown, but can in principle be formulated as functional of $Q$, which makes the calculations self-consistent. The search for $O[Q]$ has then be replaced by the search for $v_{\rm aux}[Q]$. This is sometimes equivalent to searching for an energy functional $E[Q]$ and using a variational principle.

In the present work, we will compare two cases for the choice of $Q$: the density $n$, and the one-body Green's function $G$. This means that we compare two frameworks, namely Density Functional Theory\cite{Hohenberg1964,Dreizler1990,Parr1994,Martin2004,Engel2011} (DFT) and Green's function functional theory \cite{Hedin1969,Kadanoff1962,Fetter1971,Gross1991,Stefanucci2013,Martin2016} (GFFT). These may seem to be two distinct worlds - it is often said that DFT is computationally efficient, but lacks systematic approximations\cite{Burke2012,Ruzsinszky2011}, whereas functionals of \GF s are built with systematic perturbation theory, but require more computational effort\cite{Onida2002}. Here we will explore the reasons underlying these statements, and discuss possible evolutions.

To make the discussion more focused, we will concentrate on one line of approximations. This starts with the consideration that powerful approximations for a real system may be obtained if one can profit from knowledge obtained from a simpler system, a model. This can happen in two ways: either one can directly replace the observable or auxiliary potential in the real system with the one of the model system, whose parameters are specified in an appropriate way. This is easy if $O[Q]$ or $v_{\rm aux}[Q]$ do not depend on all details of $Q$, but, for example,  only on its value in a given point $x_0$: if $O[Q]$ is in this sense \textit{near-sighted}, one can choose the parameters of the model system such that the quantity in the model $Q\mo$  has the same value in $x_0$, while the rest of $Q\mo$ can be different, and therefore simpler. Another straightforward case would be a perfectly \textit{far-sighted} situation, where $O[Q]$ depends only on an average of $Q$, and the model system can be chosen such that the average of $Q\mo$ equals the average of $Q$.

Usually, life is not that easy, but at least the appropriately specified model system may be a good starting point for a perturbation expansion. The hope would be that the resulting expansion converges rapidly. One has a tendency to consider expansions to be systematic approximations,  having in mind converging series, where approximations can be made better and better. 

\GF s functional theory is often thought to be systematic, because one associates it with Many-Body Perturbation Theory\cite{Fetter1971,Gross1991} (MBPT): this  implies a specific choice, namely, an expansion in the Coulomb interaction, starting from a model of non-interacting electrons subject to the external potential of the real system. It is mandatory to expand beyond this model, because the model cannot be made good enough to yield a satisfactory prescription of many observables that are at the center of interest of GFFT, such as spectral functions. The convergence problems of the perturbation expansion, though, make that it does not always lead to systematic improvements. 

In the framework of DFT the typical model system is the homogeneous electron gas (HEG): this model has an external potential that is very much simplified with respect to the one of the real system, but the model is fully interacting\cite{Giuliani2005}. It is mostly used to approximate directly the auxiliary potential, in many cases successfully; therefore, perturbation expansions are less prominent in the DFT framework, and more work is going into an optimized use of the HEG, which is why DFT approximations are often considered to be less systematic. 
Our goal here is to enter the discussion without prejudice, and work out why certain strategies for approximations appear naturally - or if not, maybe for historical reasons - in the two respective frameworks.

The choice of the model, or zero-order of the expansion, plays the central role. Explicitly or implicitly, it is the crucial step in all the approximations considered here. Recently, an approach termed ``connector theory'' has been introduced that aims at improving this starting point to an extent that may avoid the need for higher orders\cite{Vanzini2019}. We will further analyse and illustrate this approach, and, more generally, we will discuss the design of optimized model systems for making calculations of interacting electron systems more precise, and/or faster. While we do not claim breakthrough for a full solution of the many-body problem, we hope that this analysis and our suggestions may give inspiration for trying out a few ideas, and making collective progress.
 
 We will start by briefly reviewing DFT and GFFT in the next two sections, putting the two theories on a similar footing. Subsequently, we will discuss two aspects: first, in Section \ref{sec:balloon} we will explore how a diagrammatic expansion in terms of the density would work out, and secondly, in Section \ref{sec:connector} we will give a brief review of the connector theory. Section \ref{sec:density-matrix} will go into detail concerning much of the discussion by focussing on the density matrix, which will play the role of $O$, and for which we will search for a density functional. A brief section on time non-local observables will then lead to the final conclusions and outlook.

\section{Density Functionals}

DFT is a very general theory. In the present work, we focus on DFT for interacting electrons. Moreover, we do not consider problems linked to the spin of the electrons, and always refer to the total charge density $n(\br)$.

\subsection{The density}
In DFT, the charge density plays the role of the quantity $Q$ in terms of which functionals are built. In second quantization, at zero temperature and for fixed electron number the density is

\begin{equation}
    n(\br) = \langle N|\hat \Psi^{\dagger}(\br)\hat \Psi(\br)|N\rangle ,
    \label{eq:density}
\end{equation}
where $|N\rangle$ is the $N$-body ground state and $\hat \Psi(\br)$ is the field operator that destroys an electron in point $\br$. This expectation value can be evaluated, for example, by using Quantum Monte Carlo\cite{Foulkes2001,Kolorenc2011,Martin2016} (QMC). Here we are interested in functionals, and indeed, if one knew the total energy as functional of the density, the latter could be obtained by minimizing the energy. 

\subsection{Auxiliary system}

Kohn and Sham\cite{Kohn1965} have shown that the charge density of an interacting system can be obtained as solution of a non-interacting auxiliary system with an appropriately chosen potential, the Kohn-Sham (KS) potential. This potential can in principle be obtained from a functional derivative of the  ground state energy, 
using the variational principle. However, since the density functional of the energy is unknown, so is the Kohn-Sham potential. Moreover, the kinetic energy in the Kohn-Sham system is expressed in terms of its orbitals, because even the non-interacting kinetic energy is not known as explicit functional of the density.  

\subsection{Observables}
Only the expectation values of local one-body operators $\hat O_1$ are easily expressed as explicit density functionals, $O_1=\int d\br\,O_1(\br)n(\br)$. Other functionals are unknown and must be approximated. 

\subsection{Approximations}
A major effort is made to approximate the exchange-correlation part of the total energy as functional of the density, whereas other observables, in particular spectral functions that are also in principle functionals of the ground state density, seem to be out of reach. 
For the lack of better approximations, observables are 
 sometimes replaced by the corresponding expectation values calculated in the Kohn-Sham auxiliary system. The such obtained approximation is a density functional, and it may be considered to be an approximation to the true functional, but it is often a poor approximation, except for the Kohn-Sham density itself, which is in principle exact. 

\subsubsection{Model systems}
DFT has been successful early on because Kohn and Sham\cite{Kohn1965} suggested an efficient approximation for the auxiliary potential, the Local Density Approximation (LDA). 
This approximation is based on the hypothesis of nearsightedness\cite{Kohn1996,Prodan2005}: whereas the unknown exchange-correlation contribution  $v_{\rm xc}(\br;[n])$ depends in principle on the density in the entire system, it is dominated by the density close to the point $\br$ and can therefore be replaced by the \xc\ potential $v_{\rm xc}(\br;[n\mo])$ of a model, provided $n\mo(\br)=n(\br)$. This allows one to use as model a different HEG for each point $\br$, with the corresponding homogeneous density $n_{\br}\mo = n(\br)$.   In this way, DFT could profit from the fact that calculations in the HEG could be carried out using QMC\cite{Ceperley1980,Ortiz1994}. The importance of this idea cannot be stressed enough: the  QMC results were carried out once and forever and made available to all DFT practitioners, who therefore never had to solve the interacting electron problem. This aspect certainly explains much of the ``computational efficiency'' of DFT. 

The LDA is not exact, and the search for better approximations is ongoing. The present article is not meant to be a review paper, and we mention only the two lines of research that fit to our discussion: first, efforts have been made for an improved use of the HEG, such as the weighted density approximation or the average density approximation\cite{Gunnarsson1976a,Gunnarsson1977,Alonso1977,Alonso1978,Gunnarsson1979,Gunnarsson1980}, where the requirement for perfect nearsightedness is dropped by taking the HEG at an appropriately averaged density. Second, the most natural idea was to use the LDA as starting point for further expansions, as discussed next. 

\subsubsection{Expansions}

Starting from the homogeneous density, the logical next step is to consider density variations through gradient expansions. This leads to explicit functionals of the density and its gradients. Since a straightforward gradient expansion\cite{Hohenberg1964,Kohn1965,Ma1968,Rasolt1975} has convergence problems, successful generalized gradient approximations\cite{Langreth1980,Langreth1981,Perdew1985,Perdew1986,Perdew1986a,Perdew1996} (GGA) make use of additional knowledge, such as exact constraints, or parameters fitted to prototype systems. 

There have also been attempts to use straightforward Taylor expansion around the LDA \cite{Kohn1965,Palummo1999}. For $v_{\rm xc}$ such an expansion reads
\begin{equation}
    v_{\rm xc}(\br,[n]) = v_{\rm xc}(\br,n^h_{\br}) + \int d\brp\,\frac{\delta v_{\rm xc}(\br)}{\delta n(\brp)}_{|n=n_0}(n(\brp)-n^h_{\br}) + \ldots \,\, ,
    \label{eq:expand-vxc}
\end{equation}
where $n^h_{\br}$ is a homogeneous density distribution with value $n(\br)$. There has been discussion concerning the density distribution $n_0$ where the derivative should be taken\cite{Gunnarsson1979,Palummo1999}; strictly speaking, the Taylor expansion would prescribe $n_0=n^h_{\br}$, but other choices may improve the result when one stops at first order. No conclusive results were obtained, and the attempt was limited to an expansion of the total energy or $v_{\rm xc}$, but, as we will discuss later, the general idea of expanding in the density around an appropriate starting point may still show its power.

Also expansions in the Coulomb interaction have been proposed \cite{Goerling1994}. These do however lead to orbital dependent, not to explicit density functionals.
While being systematic, the such obtained approximations lose the extreme computational simplicity of LDA or GGAs. Finally, much effort was devoted to semiclassical expansions, see \cite{Dreizler1990}.

\section{GF functionals}

Like DFT, working with functionals of \GF s is a very general topic. Here we will focus on functionals of the one-body \GF\ for interacting electron systems, and, as above, do not consider spin explicitly. 

\subsection{The one-body Green's function}
In \GF s functional theory, most often the one-body \GF\ plays the role of $Q$. In second quantization, for a static system at zero temperature and for fixed electron number the time-ordered one-body \GF\ is defined as

\begin{equation}
    G(\br,t,\brp,t') = -i\langle N|T\left [\hat \Psi(\br,t)\hat \Psi^{\dagger}(\brp,t')\right ]|N\rangle ,
    \label{eq:greens}
\end{equation}
where $T$ indicates the time-ordering of the operators. The diagonal of this \GF , taken as $\brp=\br$ and $t'\to t^+$, equals the density \refeq{eq:density}. The density matrix is obtained in the same time limit, considering diagonal and off-diagonal elements in space. Since the external potential is supposed to be static, the \GF\ depends only on the time difference $t-t'$ or, equivalently, on one frequency $G(\br,\brp,\omega)$.  The poles of $G(\omega)$ are electron addition and removal energies. Therefore its spectral function, which is proportional to the imaginary part, is a good approximation for direct and inverse photoemission spectra\cite{Hedin1999}. The smallest differences between addition and removal poles correspond to the fundamental gaps. 

As in the case of the density, one would like to avoid calculating the \GF\ through a direct evaluation of the expectation value \refeq{eq:greens}. QMC methods are almost exclusively limited to calculations on the imaginary energy axis, and the analytic continuation to the real axis, which would yield spectra, is notoriously difficult\cite{Jarrell1996}. The way to proceed is therefore analogous to the DFT approach: one can build an energy functional\cite{Luttinger1960,Klein1961,Baym1961,Baym1962} of $G$, and the physical \GF\ is obtained at its extrema, which however, contrary to DFT, do not have to be minima.

If one evaluates \refeq{eq:greens} for the Kohn-Sham auxiliary system, one obtains
\begin{equation}
  G_{KS}(\br,t,\brp,t') = i\sum_v\phi_v(\br)\phi_v^*(\brp) e^{-i\varepsilon_v(t-t')} -i \sum_c\phi_c(\br)\phi_c^*(\brp) e^{-i\varepsilon_c(t-t')} ,
\end{equation}
where $\phi_v$ and $\varepsilon_v$ ($\phi_c$ and $\varepsilon_c$) are orbitals and eigenvalues of occupied (unoccupied) Kohn-Sham states. The \GF\ of the exact KS system is different from the exact $G$; the latter has in general a much richer spectral function, and the fundamental gaps differ because exact KS eigenvalues are not electron addition or removal energies. Nevertheless, $G_{KS}$ is often used as approximation, or as valid starting point for further calculations.

\subsection{Auxiliary system}
The variational approach to the energy as functional of the \GF\ yields the auxiliary system, similar to DFT. It can be formulated as a Dyson equation,
\begin{equation}
    G(1,2) = G_0(1,2) + G_0(1,\bar 3)\Sigma(\bar 3,\bar 4)G(\bar 4,2),
    \label{eq:dyson}
\end{equation}
where $G_0$ is the non-interacting \GF\ of the system, the compact arguments $1\equiv (\br_1,\sigma_1,t_1)$ denote space, spin and time, and the bar indicates integration: $f(\bar 1)g(\bar 1)\equiv \int d1\,f(1)g(1)$. The self-energy $\Sigma(3,4)$ consists of the Hartree potential $\delta(3-4)v_H(\br_3)$ and the remainder $\Sigma_{\rm xc}$, which contains all exchange-correlation effects and plays the role of a space, spin and time non-local auxiliary potential. When $\Sigma_{\rm xc}$ is replaced by the exchange-correlation part of the KS potential $v_{\rm xc}$, the Dyson equation yields the KS \GF .

\subsection{Observables}
Expectation values of all space local and non-local one-body operators are simple functionals of the \GF ,$$O_1=-i\int d\br d\brp\,O_1(\br,\brp)G(\br,t,\brp,t^+),$$ since its time diagonal is the one-body density matrix. Moreover, the spectral function is $$A_{\ell\ell}(\omega) = \frac{1}{\pi}|{\rm Im}\,G_{\ell\ell}(\omega)|,$$ where $\ell$ denotes matrix elements in a basis. Finally, the total energy can be expressed in terms of $G$ using the Galitzkii-Migdal expression\cite{Galitskii1958}. Therefore the main problem for many interesting applications, such as the calculation of photoemission spectra, is not to find the \GF - functional for the observable, but to find $G$ itself. Other functionals for observables related to $N$-body operators with $N\geq 2$, such as absorption spectra, have instead no simple expression in terms of the one-body $G$, and must be approximated\cite{Strinati1988,Onida2002,Bechstedt2014}. This also holds when one wishes to use variational energy functionals, which yield the same result as the Galitzkii-Migdal expression when $G$ is the solution of the Dyson equation, but may be better for an approximate $G$ that is not calculated self-consistently from the Dyson equation.

\subsection{Approximations}
To find $G$ one in general uses the Dyson \refeq{eq:dyson}, and analogous to the case of DFT, the problem is then to find good approximations for the auxiliary potential, i.e. the self-energy, as functional of $G$. By the way of contrast, there are also attempts to approximate directly the \GF , without passing through an auxiliary quantity\cite{Schirmer1983,Niessen1984}. As we will see later, this is however difficult when one is interested in the poles of $G$.

\subsubsection{Model systems}
In view of the parallels between DFT and GFFT, one might expect similar approximation strategies. In particular, since the LDA is one of the cornerstones for the success of DFT, the question is whether one can use a model system to simulate a real system also in the case of \GF s. Indeed, the closest approach to the LDA is probably dynamical mean field theory\cite{Georges1996} (DMFT) in the single-site approximation: In this approach, one is not interested in the full \GF , but only in its diagonal in a site basis, $G_{\rm \ell\ell}(\omega)$. The auxiliary system to produce this diagonal, in principle exactly, has a local but frequency-dependent auxiliary potential, i.e. a local self-energy $\tilde \Sigma_{\rm \ell\ell}(\omega)$; note that it should not be confused with the diagonal element of the full self-energy, i.e.  $\Sigma_{\rm \ell\ell}(\omega)\neq \tilde \Sigma_{\rm \ell\ell}(\omega)$ in general. As in the LDA, the hypothesis is nearsightedness, i.e. one assumes that $\tilde \Sigma_{\rm jj}(\omega)$ on a site $j$ depends only on $G_{\rm jj}(\omega)$, and not on  $G_{\rm \ell\ell}(\omega)$ on sites $\ell\neq j$ elsewhere. This allows one to import $\tilde \Sigma_{\rm jj}(\omega)$ from a model system that has the same  $G_{\rm jj}(\omega)$ on that site, and can be different elsewhere. The model chosen by single-site DMFT is the Anderson Impurity Model\cite{Anderson1961} (AIM), which can be tuned to yield the desired $G_{\rm jj}(\omega)$ by tuning a bath \GF . This leads to a self-consistent calculation of $\tilde \Sigma_{\rm \ell\ell}(\omega)$ and $G_{\rm jj}(\omega)$. The method is hence strictly analogous to the LDA, with three main differences: (i) the choice of the model system, which in the case of DMFT is meant to be closer to the target applications containing localized electrons, (ii) the fact that even in the model the interacting and non-interacting local \GF\ are different, which is not the case for the density in the HEG, and (iii) the fact that the HEG DFT results have been calculated once and forever, whereas the solutions of the AIM are not tabulated and have to be produced for each calculation. This means that DMFT does not benefit from already existing calculations of interaction effects in the model, which makes it computationally much heavier. A good reason for this is of course the fact that the HEG is parametrized by just a number, i.e. its density, whereas the AIM has a much larger parameter space (orbitals, frequency). Nevertheless, one might hope hat with increased storage capacities and machine learning interpolations, this redundancy will be removed in the future.

It is also worthwhile to mention a much simpler and approximate approach, which attempts to use the HEG as model system for \GF s. The connection between the real and the model system is made through the density; indeed, like every expectation value also the self-energy can be formulated as a functional of the density. The approximation termed Quasi-Particle Local Density Approximation\cite{Sham1966a,Wang1983} (QPLDA) is meant to describe the quasi-particle states in the spectrum close to the fermi level, and it makes again use of the hypothesis of nearsightedness. As we will discuss later, nearsightedness is easy to define (though not necessarily valid) only for an object that depends on one space coordinate, whereas the self-energy depends on $\br$ and $\brp$. Therefore, in the QPLDA approach the self-energy is first converted into an effective local potential.

The most important model for GFFT, however, is that of a system of non-interacting electrons, which gives rise to $G_0$ in the Dyson equation. 
In this model the external potential equals the one of the real system, but the interaction is put to zero. Contrary to a model such as the interacting HEG, this non-interacting model is not supposed to be tuned: it can merely be used as  starting point for further refinements which, as we will discuss in the following, imply expansions around $G_0$. One can, of course, imagine alternatives, which could take the form of a simplified but tunable interaction, instead of a rigid putting the interaction to zero, and inspirations might be obtained e.g. by work on range-separation \cite{Savin1996,Toulouse2004}. Still, there is no established method yet in this sense.

As a final remark, it should also be noted that it would probably be meaningless to tabulate the non-interacting model, since it depends on all details of the real system, through the external potential, although interpolations might be tempted.  It is, however, comparably straightforward to solve. Finally, we should add that most often the model is not a strictly non-interacting one, but rather an effective non-interacting system, such as the KS one, which has proven to be a good starting point for further developments of GFFT.

\subsubsection{Expansions}

Starting from the non-interacting model the most natural thing to do is an expansion in the Coulomb interaction  $v_c(|\br-\brp|)$ around $v_c=0$. 
This is the heart of many-body perturbation theory\cite{Fetter1971,Mahan1981}. It gives rise to the diagrammatic expansions in terms of $v_c$ and $G_0$ that are schematically
$G = G_0 + \frac{\delta G}{\delta v_c}_{|v_c=0}v_c + \ldots$.
As we will see later, it is difficult to use expansions when one is interested in functions with poles. Therefore, except for some quantum chemistry work (see e.g. \cite{Schirmer1983,Niessen1984,Holleboom1990}),  rarely the \GF\ itself is expanded, and one rather seeks to approximate the self-energy, i.e. the auxiliary system, as in DFT. This pushes the problem to the next level, because also the self-energy has poles that suffer from the approximations, but the resulting spectra are overall more decent than those resulting from a direct expansion of the \GF . One may note that other analogues of an auxiliary system are possible, in particular the cumulant representation\cite{Aryasetiawan1996,Mahan1981} of the \GF , where one approximates the cumulant $C$ in $G\equiv G_0e^C$.

Comparison with DFT expansions in terms of the interaction \cite{Goerling1994} would be an interesting subject on its own. However, in the spirit of standard DFT approximations, and keeping in mind that the realistic possibility to tabulate a model system gives a huge advantage to such an approximation, in the following we will rather focus on expansions around an interacting model system with simplified potential. 

To expand around an interacting model system, we have several possibilities, because the self-energy can be treated as functional of the potential, of the interacting or non-interacting \GF , or even of the density. Most naturally, we would consider $\Sigma_{\rm xc}$ as functional of $G$, which would yield
\begin{equation}
    \Sigma_{\rm xc}(1,2) = \Sigma_{\rm xc}\mo(1,2) + \frac{\delta \Sigma_{\rm xc}(1,2)}{\delta G(\bar 3,\bar 4)}_{|G=G\mo}(G(\bar 3,\bar 4)-G\mo(\bar 3,\bar 4)) + \ldots
\end{equation}
Even in a model, even the first derivative will in general have to be approximated, for example, using the GW approximation\cite{Hedin1965} $\Sigma_{\rm xc}(1,2)=iG(1,2)W(12)$ for the self-energy, where $W$ is the screened interaction. If one further neglects the derivative of $W$, the result is $\Sigma_{\rm xc} \approx \Sigma_{\rm xc}\mo - iW\mo G\mo + iW\mo G$: it consists of all diagrams in the model, except for the GW diagram that is partially evaluated in the real system. For more advanced self-energies in the derivative, calculations become cumbersome: although the derivative $\Xi\mo$ is calculated in the model and might therefore be tabulated, to evaluate $\Xi\mo(1,2;\bar 3,\bar 4)G(\bar 3,\bar 4)$ would require more computational effort than the GW approximation. The usefulness of this expansion is therefore not established, although it might be worthwhile to explore it further.

\hspace{3cm}

\section{Diagrammatic expansions for explicit density functionals}
\label{sec:balloon}

In spite of all possible shortcomings, expansions remain one of the most promising ways to improve upon a given approximation. After our brief discussion of  expansions in terms of the Coulomb interaction or an interacting \GF , we will dedicate the present section to expansions in terms of the density.

\subsection{Motivation}

We postulate that one very strong criterion for the choice of an approximation strategy is our ability to avoid redundancy: the idea is to profit from already existing calculations, which therefore have to be carried out only once and forever, serving numerous different applications. As discussed above, this is most easily achieved by using an interacting model system with a simple, but tunable potential. Such a model can be considered as the starting point around which we will expand. The most straightforward move would be to expand directly in the difference between the real and the model potentials, leading to corrections that are linear and higher order response. Alternatively, thanks to DFT we can expand in the difference of real and model densities. This option may have a major advantage: the density is the diagonal of the one-body \GF , and as we have seen, $G$ gives direct access to several interesting observables, and to others, though less immediately, through MBPT. With the tight relation between the \GF\ and the density, there is hope that one can have error cancelling, and/or use exact constraints to improve results. 

Expanding in the density around a model system will yield observables or auxiliary potentials as explicit functionals of the the model density and of variations around it. These functionals will consist of universal building blocks stemming from the model, which are calculated once and forever, and simple integrals to be performed with the density variations.
Here ``universal'' means that these building blocks do not depend on the real system of interest, within the class of real systems that may be described starting from a given model. Of course,  the ambition is to find models that are general enough to serve for a huge class of real systems. The HEG plays a special role here, since it is probably the most widely used model in electronic structure calculations; it has also be shown that it can be considered to be the first step in an expansion in $\hslash$ \cite{Elliott2008}. However, with the advent of computer power and storage capacities, it will be most interesting to examine also alternative models, which may have more degrees of freedom.

\subsection{Expansion in the density around a model system}

The strategy is to express an object $O[n]$ in a functional Taylor expansion around $O[n\mo]$.
To be general enough, we suppose that $O$ may be non-local in space and time, which includes the one-body \GF . The expansion reads 
\begin{eqnarray}
    &&O(\br,\brp,t-t';[n]) = O (\br,\brp,t-t';[n\mo])\nonumber\\
   & &+
    \int d\brpp\, \frac{\delta O(\br,\brp,t-t';[\tilde n])}{\delta \tilde n(\brpp)}_{|\tilde n = n\mo}\Delta n(\brpp)\nonumber\\
    &+& \frac{1}{2}\int d\brpp d\brppp\,\frac{\delta^2 O(\br,\brp,t-t';[\tilde n])}{\delta \tilde n(\brpp)\delta \tilde n(\brppp)}_{|\tilde n = n\mo}\Delta n(\brpp)\Delta n(\brppp)\nonumber\\
    &+& \ldots  
    \label{eq:expansion-o-n-int}
\end{eqnarray}
where the density difference $\Delta n$ is defined as
\begin{equation}
    \Delta n(\br)\equiv n(\br)-n\mo(\br).
\end{equation}
We now insert the external potential $v\ex$ in a chain rule\cite{Leeuwen2001,Leeuwen2003}. 
This is interesting for later discussions, and it corresponds to a choice that one would often make in practice.
The chain rule yields:
\begin{eqnarray}
&&O(\br,\brp,t-t';[n]) = O(\br,\brp,t-t';[n\mo])\nonumber\\
   & &+
   \int d\brpp d\br_1\,\frac{\delta O(\br,\brp,t-t';[\tilde v_{\rm ext}])}{\delta \tilde v_{\rm ext}(\br_1)}_{|\tilde v_{\rm ext} = v_{\rm ext}[n\mo]} (\chi\mo)^{-1}(\br_1;\brpp)\Delta n(\brpp)\nonumber\\
        &&+ \ldots
\end{eqnarray}
where
\begin{equation}
    (\chi\mo)^{-1}(\br_1;\brpp)\equiv\frac{\delta v\ex(\br_1;[\tilde n])}{\delta \tilde n(\brpp)}_{|\tilde n = n\mo}
\end{equation}
is the inverse of the static response function of the model system.
In the next terms, higher order response functions appear, schematically:
\begin{eqnarray}
    \frac{\delta^2 O}{\delta n \delta n} &=& \frac{\delta^2 O}{\delta v\ex \delta v\ex} (\chi\mo)^{-1}(\chi\mo)^{-1} \nonumber\\
    &-& \frac{\delta O}{\delta v\ex}(\chi\mo)^{-1} {\chi^{(2){\rm m}}} (\chi\mo)^{-1}(\chi\mo)^{-1},
    \label{eq:expansion-o2-n}
\end{eqnarray}
where the second-order response function is
\begin{equation}
    \chi^{(2){\rm m}}(\br;\br_1,\br_2) \equiv
  \frac{\delta^2  n(\br;[\tilde v\ex])}{\delta v\ex(\br_1)\delta v\ex(\br_2)}_{|\tilde v\ex = v\ex[n\mo]}.
\end{equation}
Note that all response functions are static. $O$ depends only on a time \textit{difference}, and is fully described by a static ground state, because the system time-independent. Since there is a one-to-one relation between the interacting density, the external potential and the non-interacting density, the expansion can be made in either the interacting or the non-interacting density.

\subsubsection{Expansion in the non-interacting density}
Let us start with the expansion in the non-interacting density. We can use the above equations, interpreting the densities as non-interacting ones in the real and the model system, $n\to n_0$ and $n\mo \to n_0\mo$, and all model response functions as non-interacting  model response functions, $\chi\mo\to \chi_0\mo$.  The latter can be expressed in terms of non-interacting \GF  
\begin{eqnarray}
\chi_0\mo(\br,\brp) &=& -i\int d\tau \,G_0\mo(\br,\brp,\tau)G_0\mo(\brp,\br;-\tau) \nonumber\\
\chi_0^{(2){\rm m}}(\br;\br_1,\br_2) &=& -2i\int d\tau_1 d\tau_2\times\nonumber\\ &\times&G_0\mo(\br,\br_2,\tau_2)G_0\mo(\br_2,\br_1;\tau_1-\tau_2)G_0\mo(\br_1,\br,-\tau_1)\nonumber\\
\ldots \,\,\,\,\,,
\end{eqnarray}
where $\tau$ indicates time differences. The order $N$ response function carries a prefactor $N!$.

\begin{figure}[h]
\centering
  \includegraphics[height=6cm]{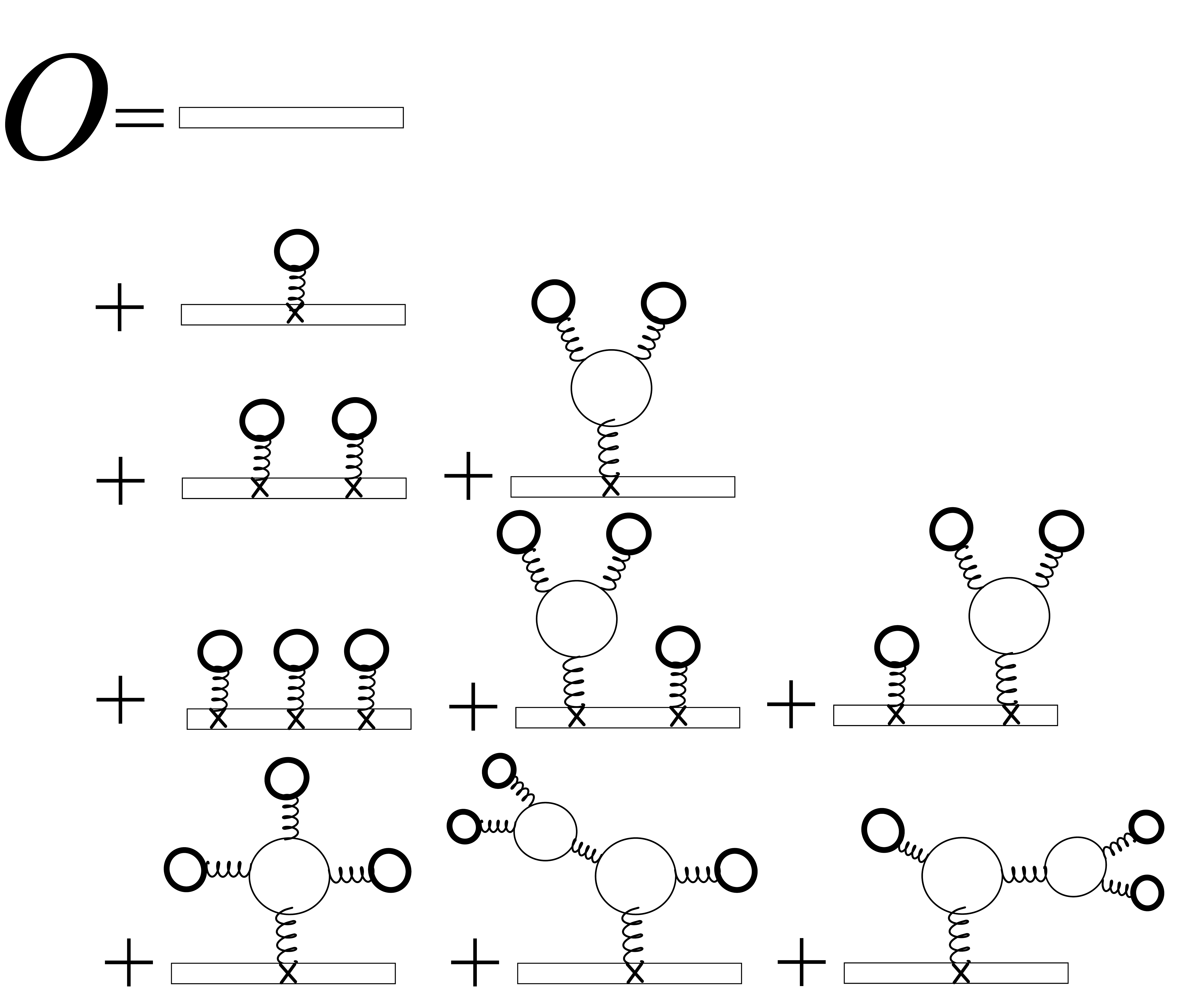}
  \caption{``Balloon'' expansion of an observable ${O}$ in the non-interacting density. ${O}[n\mo]$ its derivatives with respect to the external potential are shown by the horizontal bars. Small fat circles represent the non-interacting density $n_0$. The wiggly line is an inverse response function, and thin circles are higher order response functions. All quantities except for $n_0$ are calculated in the model.  }
  \label{fgr:balloon-expansion}
\end{figure}

The expansion can be most conveniently depicted by diagrams, as indicated in Fig. \ref{fgr:balloon-expansion}. It contains the linear response term
\begin{equation}
 \mathcal{L}(\br)\equiv \int d\brp\,(\chi_0\mo)^{-1}(\br,\brp)\Delta n_0(\brp),    
\end{equation}
which we depict as
a fat circle ($\Delta n_0$) attached to a wiggly line for $(\chi_0\mo)^{-1}$ (a ``balloon''), and larger balloons consisting of a loop built with several Green's functions $G_0\mo$ (thin lines) that is also attached to a $(\chi_0\mo)^{-1}$. Balloons can be attached to other balloons, as shown in Fig. \ref{fgr:balloon-expansion} for the first three orders.
The bar at the bottom represents $O[n\mo]$ or, when balloons are attached to it,  its derivatives.

The diagrammatic rules to build the series are the following:
\begin{itemize}
    \item Build a small balloon $\mathcal{L}$ and large balloons $(\chi_0\mo)^{-1}{G_0\mo}{G_0\mo}{G_0\mo} \ldots$
    \item Attach small or large balloons to a $G_0\mo$, or to $O\mo$.
    \item Large balloons are always dressed by at least two other balloons that can be large and/or small.
    \item Each large balloon contains a response function beyond linear response;  it carries a factor $i$, and a prefactor $N!$ corresponding to the order of the response function.
    \item  To order $O(N)$, one has a total of $N$ small balloons, and one has to add an overall prefactor $1/N!$.
\end{itemize}

If this expansion in the non-interacting $n_0$ is performed for an interacting system, all the interaction effects are contained in $O[n\mo]$ and its derivatives with respect to the external potential. These derivatives are calculated once and forever in the model, as well as all response functions (which remain non-interacting), whereas all the information about the particular system is contained in the non-interacting density. This is a way to separate interaction effects from the specification of a particular system.
As can be seen from Fig. \ref{fgr:balloon-expansion}, once the model is tabulated the main workload consists in evaluating the integrals that attach balloons to each other or to $O^m$. The difficulty of the initial step - the evaluation of the model - depends on the object $O$, for which one has to be able to calculate derivatives with respect to the external potential in an interacting model system.

\subsubsection{Expansion in the interacting density}

\begin{figure}[h]
\centering
  \includegraphics[height=6cm]{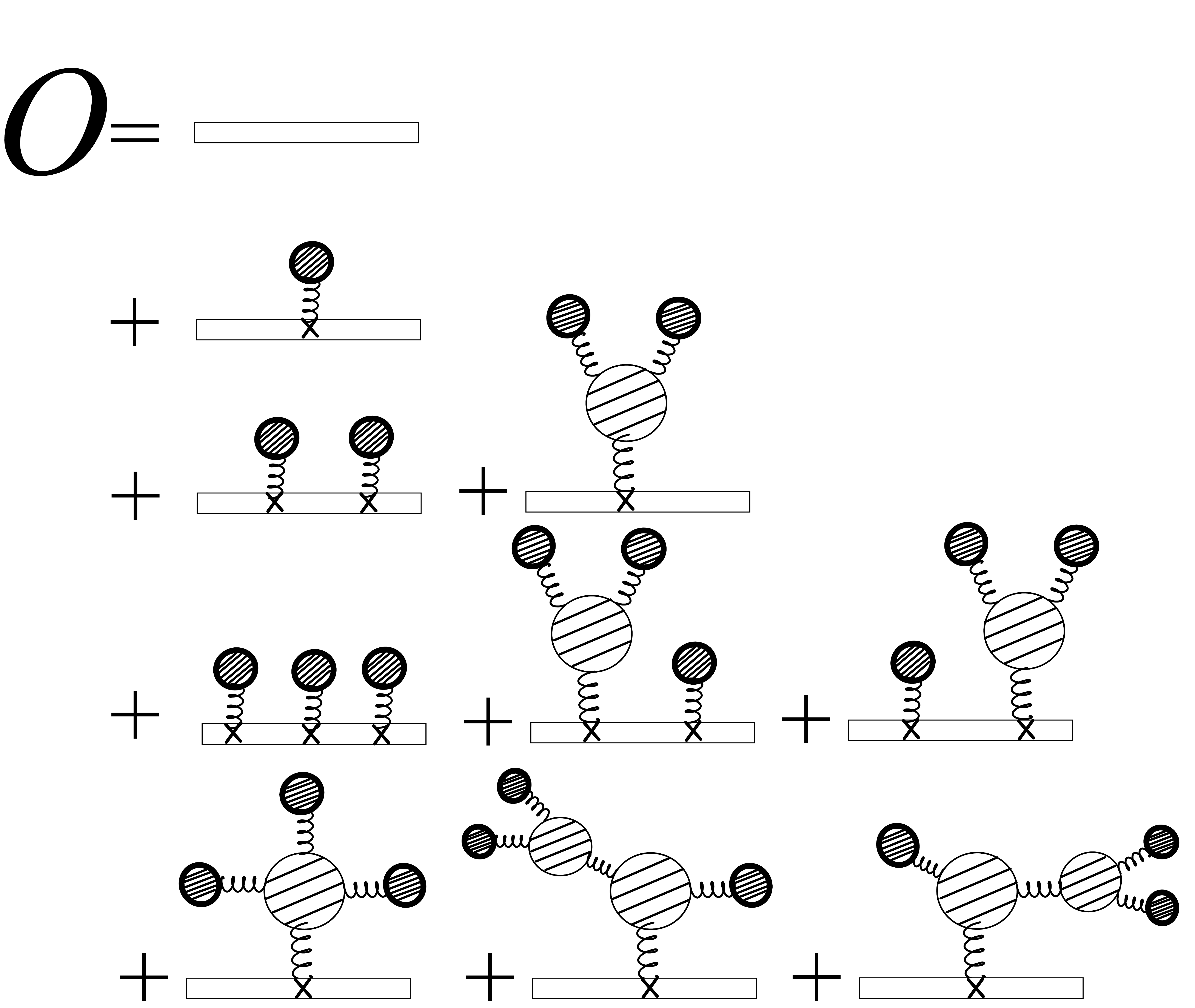}
  \caption{Balloon expansion in the interacting density, represented by filled small circles. As Fig. \ref{fgr:balloon-expansion}, but all direct and inverse response functions are interacting.}
  \label{fgr:balloon-expansion-int}
\end{figure}

To expand in the interacting density, one has to replace  $\chi_0\mo$ by $\chi\mo$ and $n_0\mo$ by $n\mo$. The resulting expansion is depicted in Fig. \ref{fgr:balloon-expansion-int}. Now fat circles stand for the \textit{interacting} density difference and wiggly lines for $(\chi\mo)^{-1}$. Large balloons contain higher order \textit{interacting} model response functions, which are depicted by a filled area, since they are no longer simple integrals of products of $G_0\mo$ nor of $G\mo$. The bottom bar represents $O\mo$ and its derivatives, as before.  

The expansion in the interacting density does not separate interaction and system effects: $O\mo$ and its derivatives as well as the response functions  contain only interaction effects, but the density difference contains both system and interaction effects, which makes its calculation not trivial. For the evaluation of some observables $O$ it may be sufficient to use Kohn-Sham DFT with an approximate functional; if $O$ is the KS $v_{\rm xc}$ itself, the calculation may be done self-consistently. 

In the chain rule, one can also use the full KS potential instead of the external one. All direct and inverse response functions are then non-interacting KS ones. When the model is chosen to be the HEG, the Hartree and {\xc} potentials do not change the response functions. In this case, the effect of the Coulomb interaction are contained only in $\Delta n$ and in $O\mo$ and its derivatives with respect to the KS potential, and, in case $O=G_{KS}$, exclusively in $\Delta n$.

\hspace{2cm}

One may wonder why the expansion in $n$ should be better than the expansion in $n_0$, which is appealing because of the separation of system- and interaction- effects. While it is difficult to make a general point, some insight can be gained when we choose $O$ to be the \GF . Since $G$ gives direct access to many interesting observables, it is also a key example. 

\subsection{Expansion of the one-body \GF\ }

When $O$ in Eq. \ref{eq:expansion-o-n-int} or in its interacting counterpart is the one-body $G$,  the derivatives of $G$ make generalized response functions appear, with a non-locality as indicated by the pre-superscript with reference to the number of space arguments: 
\begin{eqnarray}
    &&{^3\chi\mo}(\br,\brp,t-t';\brpp) =  -i \frac{\delta G(\br,\brp,t-t';[\tilde v\ex])}{\delta \tilde v\ex(\brpp)}_{|\tilde v\ex = v\ex\mo}\nonumber\\
    && {^4\chi^{(2)}{}\mo}(\br,\brp,t-t';\brpp,\brppp)  = -i \frac{\delta^2 G(\br,\brp,t-t';[\tilde v\ex])}{\delta \tilde v\ex(\brpp)\delta \tilde v\ex(\brppp)}_{|\tilde v\ex = v\ex\mo}\nonumber\\
    && \ldots \,\,\,\,\,\, .
    \label{eq:derivatives-G-int}
\end{eqnarray}
The whole expansion is hence a chain of direct and inverse response functions.
This allows one to make a quick consistency check: the density resulting from this expansion is the diagonal of the final \GF . This transforms the generalized non-local into an ordinary response function; graphically, it consists in closing the bottom bar in Fig. \ref{fgr:balloon-expansion-int} into a filled area. Taking into account the prefactors as explained above, together with the fact that $\chi\mo(\chi\mo)^{-1}=1$, one can see that the two terms in the third line cancel, as well as each term in the forth line with the term below it in the fifth line. The same holds for higher orders. The final result is the sum of the first two terms, which yield consistently the exact density.
This consistency is an important feature of the expansion; it means for example that if one is interested in $O$ the density matrix, its diagonal is automatically fixed to the exact result, to any order of the expansion.
This is instead  not true when one expands in the non-interacting $n_0$, unless $O$ is the \textit{non-interacting} \GF . It is a strong argument to expand in a density that is consistent with the object that is to be expanded. It also gives an argument in favour of expanding in the density instead of expanding in the external potential.

The result of these expansions are \textit{explicit} functionals of the model density and of the density difference, but not of $n\mo+\Delta n$. If $n\mo$ is fixed and $\Delta n$ freely variable, this is equivalent to a density functional. If model density and density difference are set up as complementary components of the density, this is instead not true for a finite-order expansion.

\subsection{Other expansion schemes}

One can obtain the balloon expansion also by iterating the Dyson equation twice. Let us use for simplicity the expansion of the non-interacting \GF , and expand the Dyson equation: 
\begin{equation}
    G_0 = G_0\mo + G_0\mo v_{\rm ext}G_0\mo + G_0\mo v_{\rm ext}G_0\mo v_{\rm ext}G_0\mo+ \ldots \,\, ,
\end{equation}
where $v\ex$ is the difference between the real and the model potential. 
Taking the diagonal, we obtain 
\begin{equation}
    v\ex = (\chi_0\mo)^{-1}\Big (n_0-n_0\mo +i \langle G_0\mo v_{\rm ext}G_0\mo v_{\rm ext}G_0\mo \rangle - \ldots \Big ),
\end{equation}
and iteration leads to
\begin{eqnarray}
{\rm O(1):}&v_{\rm ext,(1)} =& (\chi_0\mo)^{-1} (n_0-n_0\mo)=\mathcal{L}\nonumber\\
{\rm O(2):}&v_{\rm ext,(2)} =& i(\chi_0\mo)^{-1} \langle G_0\mo \mathcal{L}G_0\mo\mathcal{L}G_0\mo\rangle\,\, \ldots
\end{eqnarray}
For the \GF\ order by order in the potential, we have
\begin{eqnarray}
{\rm O(0):}&G_{0,(0)} =& G_0\mo\nonumber\\
{\rm O(1):}&G_{0,(1)} =& G_0\mo v_{\rm ext,(1)}G_0\mo\nonumber\\
{\rm O(2):}&G_{0,(2)} =& G_0\mo v_{\rm ext,(1)}G_0\mo v_{\rm ext,(1)}G_0\mo + G_0\mo v_{\rm ext,(2)}G_0\mo\nonumber\\
\ldots \,\,& \ldots \,\,& \ldots
\end{eqnarray}

To infinite order, this is exactly the balloon expansion for the non-interacting \GF , Fig. \ref{fgr:balloon-expansion}, with $\mathcal{L}$ the small balloons. One can now identify terms; for example, the first second-order term corresponds to the first second-order diagram of Fig. \ref{fgr:balloon-expansion}. 
However, while the balloon expansion is order by order in the density, here one could stop the series at a given order in $v\ex$, itself calculated to a given order in $\Delta n_0$.  The resulting density is not necessarily consistent. For example, a first-order $v_{\rm ext,(1)}$
used to build $G_0$ to second order misses the second of the second-order terms, which is necessary to cancel the spurious second-order contribution which introduces an error in the density that is of second order in $\Delta n$. This is important, since one might be tempted to use the exact relation between external potential and non-interacting \GF , $G_0=((G_0\mo )^{-1}- v\ex [n])^{-1}$, to build a density functional. This inconsistency  is a general property of Dyson equations when their kernel is determined perturbatively; in some cases, e.g. the calculation of quasi-particle energies, solving the Dyson equation improves results, whereas in other cases, e.g. satellites, it deteriorates results\cite{Blomberg1972,Guzzo2011, Bechstedt2014}. This has to be investigated on a case-by-case basis. 

There are many ways to iterate the Dyson equation explicitly or implicitly. For example, the exact KS potential for small systems has been obtained through numerical iteration \cite{Almbladh1984,Goerling1992,Wang1993,Leeuwen1994,Zhao1994,Peirs2003},
such as
\begin{equation}
 v_{\rm ext}^{(i)}(\br) = \frac{n^{(i-1)}(\br)}{n(\br)}v_{\rm ext}^{(i-1)}(\br).
\end{equation}
The same procedure can be used to write the KS potential as functional of the density: for example starting with  a HEG $v_{\rm KS}^{(0)}(\br)=\mu$, 
one  gets
$
 v_{\rm KS}^{(1)}(\br) = \mu \,n^{0}/n(\br)
$,
then $n^{(2)}=-i[(G_0^{\rm HEG})^{-1}-v_{\rm KS}^{(1)}]^{-1}(\br,\br,t,t^+)$, and so on.
However, the resulting functionals imply inversions leading to \GF s, and they would therefore be computationally less efficient if used straightforwardly, calling for additional approximations for the evaluation of the \GF .\cite{Yang1988a}

\section{Connector Theory approach}
\label{sec:connector}

The model system deserves particular attention in the present work. 
A good choice optimizes the starting point of an expansion and improves its convergence. If the model is flexible enough, it might even be possible to tune it such that it yields directly the desired result of the real system, without expanding beyond the zero-order. This is the hypothesis underlying \textit{connector theory}\cite{Vanzini2019} (CT), which we will briefly outline in the following, limiting ourselves to aspects that are important for the present work; a more complete description can be found in Ref. \cite{Vanzini2019}.

The aim is to tune the model system such that $O[Q]=O[Q\mo]$, where $Q$ can be the external potential, interacting or non-interacting density or \GF , etc. If the model is flexible enough such that the equation can be satisfied in principle, one can try find the one or more $Q\mo$ for which the equality holds. 
Of course, in order to find the exact $Q\mo$ one would have to know the final solution. However, as explained in Ref. \cite{Vanzini2019} good approximations can be obtained by using the same approximation on $O[Q]$ and $O[Q\mo]$, solving $O_{\rm approx}[Q]=O_{\rm approx}[Q^c]$ for $Q^c$. The final result is obtained as
$O[Q]\approx O[Q^c]$; the final step is for free, since for the set of model $Q\mo$s the results are tabulated. 

Expansions are good candidates for such approximations. In particular, if $Q$ depends on a set of coordinates $x$, $Q(x)$, and if $Q\mo$ is instead simply one number, the first order \textit{connector} $Q^c$ reads
\begin{equation}
Q^c= \frac{\int dx'\, f(x')Q(x')}{\int dx'\,f(x')},
\end{equation}
where $f(x')\equiv (\delta O/\delta Q(x'))_{|Q=Q\mo}$. 
If this first order approximation to the connector is meaningful, $f(x')$ is the quantity that determines the range of $x'$ that is important to describe $O$. Therefore, when $O$ itself depends explicitly on $x$, i.e. $O=O(x,[Q])$, the derivative
$f(x,x')\equiv (\delta O(x)/\delta Q(x'))_{|Q=Q\mo}$ 
 is the quantity that determines the far- or near-sightedness of $O$. Using the connector avoids then the necessity to make a guess.
 
 When $Q$ is the density, $Q^c[Q]$ is a functional of the density, and so is the final $O[Q^c]$. Connector theory in that case gives us new density functionals.

The workload to evaluate this first-order connector is determined by the calculation of the first order. However, using the connector can improve the result with respect to that of the first-order expansion itself, as we will illustrate.

\section{Spatially non-local observables: the density matrix}
\label{sec:density-matrix}

To become concrete, we will study the one-body density matrix (DM), which is linked to the \GF\ as $n(\br,\brp) = -iG(\br,t,\brp,t^+)$. The DM is a fundamental ingredient for the understanding and calculations of many-electron systems\cite{Coleman1963,Levy1979,Pernal2016}: for example, it gives direct access to the kinetic and exchange energies, and the occupation numbers, which are its eigenvalues, indicate the degree of correlation of a system.  At the same time, it is non-local, in the sense that it depends on two space arguments: it is therefore not straightforward\cite{Kohn1965,Sham1973,Gunnarsson1979,Engel1995,Palummo1999,Patrick2015} to propose an approximation such as the LDA, since one has to define nearsightedness in a more general way. 

\subsection{Illustration: the single-electron density matrix}

In order to get more insight, it is useful to look at a situation where analytically exact results are available. The simplest case, independently of the external potential, is that of one single electron.

\subsubsection{Exact relations}
For one electron,  the occupied orbital is the square root of the density, and the exact density matrix reads
\begin{equation}
    n({\bf r}_1,{\bf r}_2) = \sqrt{n({\bf r}_1)n({\bf r}_2)}.
    \label{eq:sqrt-exact}
\end{equation}
This DM depends only on the densities in $\br_1$ and $\br_2$: this is a generalized nearsightedness for an object evaluated in points $(\br_1,\br_2)$.

Starting from the one-electron Schr\"odinger equation for the occupied orbital,
\begin{equation}
    -\frac{\nabla^2}{2} \sqrt{n(\br)} + v\ex (\br)\sqrt{n(\br)} = \varepsilon \sqrt{n(\br)}.
\end{equation}
one obtains the external potential, a part from the constant $\varepsilon$,
\begin{equation}
    v\ex (\br;[n]) = \frac{1}{\sqrt{n(\br)}}\frac{\nabla^2}{2} \sqrt{n(\br)}  + \varepsilon :
    \label{eq:pot-nfctl-exact}
\end{equation}
the potential is not perfectly nearsighted in the density, but depends locally on the density and its gradients.

For the single electron the DM is a frequency integral of the \GF\ 
\begin{equation}
    G_0(\br,\brp;\omega) = \left(\omega +\frac{\nabla^2}{2}-v\ex -i\eta\right)^{-1}(\br,\brp); 
    \label{eq:gf0}
\end{equation}

it depends on the external potential in a non-local way, and it would be difficult to use nearsightedness to build a potential functional. 
Let us now consider the exact expressions as 
benchmark for approximations based on expansions. 

\subsubsection{Expansions}
To get some insight, let us first suppose that we use a Taylor expansion around a model system to approximate the DM.

\paragraph{First order expansion in the density.}
A first order expansion of \eqref{eq:sqrt-exact} around a model density yields
\begin{eqnarray}
n(\br_1,\br_2) \approx n\mo(\br_1,\br_2) \Big [ 1 + \frac{\Delta n(\br_2)}{2n\mo(\br_2)}+\frac{\Delta n(\br_1)}{2n\mo(\br_1)}\Big ].
\label{eq:exp-gamma-o1}
\end{eqnarray}
The same expression can be obtained from  the balloon expansion, Sec. \ref{sec:balloon}, for the time-diagonal of the non-interacting \GF .
The static direct and inverse response functions of the model  in terms of the model one-electron eigenvalues $\varepsilon_n$ and wavefunctions $\phi_n$ are:
\begin{equation}
    \chi_0\mo(\br_1,\br_2) = 2\sum_c\frac{\phi_c(\br_1)\phi_c(\br_2)}{\varepsilon_v-\varepsilon_c}\phi_v(\br_1)\phi_v(\br_2),
    \label{eq:chi0-sum}
\end{equation}
where $v$ ($c$) are the only occupied state (empty states). For simplicity, all wavefunctions are taken to be real. The inverse response function is\footnote{As can be seen from \eqref{eq:direct3-inverse}, strictly speaking this inverse is  missing a contribution, which does however not contribute to the first order result \eqref{eq:o1-final}. In other words, it does not change $(\chi_0\mo)^{-1}\Delta n$ as long as the number of electrons remains unchanged between the model and the real system. }
\begin{equation}
    (\chi_0\mo)^{-1}(\br_3,\br_1) = \frac{1}{2\phi_v(\br_3)\phi_v(\br_1)}\sum_{c}\phi_c(\br_3)\phi_c(\br_1)(\varepsilon_v-\varepsilon_c),
    \label{eq:chi0-1-m}
\end{equation}
and the derivative of the density matrix,  
\begin{eqnarray}
 {^3\chi}_0\mo(\br_1,\br_2;\br_3) &=& \sum_c\frac{\phi_c(\br_3)\phi_c(\br_2)}{\varepsilon_v-\varepsilon_c}\phi_v(\br_1)\phi_v(\br_3)\nonumber\\
 &+& \sum_c\frac{\phi_c(\br_1)\phi_c(\br_3)}{\varepsilon_v-\varepsilon_c}\phi_v(\br_3)\phi_v(\br_2).
 \label{eq:3chi0-m}
\end{eqnarray}
For the first order correction we evaluate
\begin{multline}
{^3\chi}_0\mo(\br_1,\br_2;\bar \br_3)(\chi_0\mo)^{-1}(\bar \br_3,\br_4)= \frac{\phi(\br_1)}{2\phi(\br_4)}\delta(\br_2-\br_4) \\
+
\frac{\phi(\br_2)}{2\phi(\br_4)}\delta(\br_1-\br_4) -n\mo(\br_1,\br_2),
\label{eq:direct3-inverse}
\end{multline}
where the orthonormality of orbitals and the completeness relation were used, and integration with $\Delta n(\bar \br_4)$ yields 
\begin{equation}
    n(\br_1,\br_2) = n\mo (\br_1,\br_2) + \frac{\phi(\br_1)}{2\phi(\br_2)}\Delta n(\br_2) + \frac{\phi(\br_2)}{2\phi(\br_1)}\Delta n(\br_1),
    \label{eq:o1-final}
\end{equation}
because $\int d\bar \br_4 \Delta n(\bar \br_4)=0$.
This equals therefore
\eqref{eq:exp-gamma-o1}, 
the expansion of the exact result \eqref{eq:sqrt-exact} to first order in $\Delta n$.

One might think that the nearsightedness is a direct consequence of the short range of the response. 
However, let us examine ${^3\chi_0}$. For the one-electron case, we have 
\begin{equation}
    {^3\chi_0}(\br_1,\br_2;\br) = \frac{\delta n(\br_1,\br_2)}{\delta v\ex(\br)} = 
    \frac{\sqrt{n(\br_2)}}{2\sqrt{n(\br_1)}}\chi_0(\br_1,\br) + 
     \frac{\sqrt{n(\br_1)}}{2\sqrt{n(\br_2)}}\chi_0(\br_2,\br),
     \label{eq:form}
\end{equation}
which immediately leads to the generalized nearsightedness, because $\chi_0\chi_0^{-1}$ is always short ranged. Note that the response function alone could even be very long ranged.
Of course, if ${^3\chi_0}$  is not of form (\ref{eq:form}) but the response functions are (generalized) short-ranged, the resulting DM is also generalized nearsighted. It remains to be seen what is dominating the nearsightedness range in systems of more than one electron.
Note that here ``nearsightedness'' only applies to the difference $\Delta n$, not to the density itself; this is perfectly suitable for our purpose of finding approximations \textit{given the results of the model system}, but different from the more general question of the behaviour of the system. If the model is the HEG, then ``nearsightedness'' refers to the nearsightedness with respect to density variations, which is closer to the general concept. 

Here we have examined only the linear response. 
From the direct expansion of \eqref{eq:sqrt-exact} and from the balloon expansion we can, however, expect that the structure of the problem does not change at higher orders, and similar arguments should apply.

\paragraph{Choice of the zero order.} 
The result of Eq. \ref{eq:exp-gamma-o1}  and higher orders depend on the choice of the model density $n\mo(\br)$, to be precise, on   $\Delta n(\br_1)/n\mo(\br_1)$ and $\Delta n(\br_2)/n\mo(\br_2)$. These ratios must be small for the expansion to converge. 
This can be used as guideline to choose the zero-order of the expansion.

The simplest choice would be to start the expansion with one and the same HEG to approximate all elements of the DM, $n\mo(\br_1)=n\mo(\br_2)=n\mo$, for example, the average density of the system. However, whereas this might be good if the density is quasi-homogeneous, in strongly inhomogeneous systems one could have $\Delta n(\br_1)/n\mo \geq 1$, which would lead to divergence of the series.
Instead of the average density, one could take the average between the highest and lowest occurring density, $(n_{\rm max}+n_{\rm min})/2$. In that case $\Delta n(\br_1)/n\mo \leq 1$ everywhere. Still, the series will converge less well when the density is close to the maximum or minimum in points $\br_1$ and/or $\br_2$, and extremely slowly when the minimum density is zero.
So, such an expansion would work straightforwardly for systems where the density variation has a small amplitude compared to the average density. 

The choice of the HEG as model system can be optimized further by allowing a different HEG for each pair $(\br_1,\br_2)$. The most natural choice would be a homogeneous density defined as $n\mo_{\br_1\br_2}(\br) = (n(\br_1)+n(\br_2))/2$, independent of $\br$ but different for each pair $(\br_1,\br_2)$. In that case, $\Delta n(\br_1)/n\mo(\br_1) = (n(\br_1)-n(\br_2))/(n(\br_1)+n(\br_2))\leq 1$, as desired, and similarly for $\br_2$. Again, however, one would expect bad convergence when one of the densities approaches zero. 
Note that taking $n(\frac{\br_1+\br_2}{2})$ would be meaningless.

To go further, let us make the hypothesis that even beyond  a single electron the density matrix is generalized  nearsighted, 
in the sense that for a matrix element at given $(\br_1,\br_2)$ only the density near $\br_1$ and $\br_2$ is important. Supposing, as in the LDA, that close to those points the density is slowly varying, one could build an \textit{inhomogeneous} model, with the requirement that $(n(\br_i)-n_{\br_1\br_2}\mo(\br_i))/n_{\br_1\br_2}\mo(\br_i) \ll 1$ for $i=1,2$.
The density $n_{\br_1\br_2}\mo(\br)$ of this model would in general be different for every $(\br_1,\br_2)$. Although inhomogeneous, such a model should 
still be simple enough to be solved with advanced methods for a whole series of its parameters. An example could be crystals with only one fourier component. 
Such an enterprise would not have been thinkable when DFT was born, but is today within reach; we will come back to this in the outlook. 

\paragraph{Higher orders.}
If the model is homogeneous, for whatsoever  $n^m$ the first order  \refeq{eq:exp-gamma-o1} yields
\begin{equation}
n(\br_1,\br_2) =  \frac{ n(\br_1)+n(\br_2)}{2}.
\label{eq:exp-gamma-o1-homo}
\end{equation}
The second order, instead, depends on $n^m$, yielding : 
$$n(\br_1,\br_2) \approx  \frac{ n(\br_1)+n(\br_2)}{2} - \frac{ (n(\br_1)-n(\br_2))^2}{8 n^m}. $$ 
If $n_{\br_1\br_2}\mo= (n(\br_1)+n(\br_2))/2$,
\begin{equation}
    n(\br_1,\br_2) \approx  \frac{ n(\br_1)+n(\br_2)}{4} + \frac{n(\br_1)n(\br_2)}{n(\br_1)+n(\br_2)}.
\label{eq:exp-gamma-o2-homo}
\end{equation} 
For a general non-homogeneous model the result reads:
\begin{multline}
n(\br_1,\br_2) \approx n\mo(\br_1,\br_2)\left(1+\frac{ \Delta n(\br_1)}{2n\mo(\br_1)} + \frac{ \Delta n(\br_2)}{2n\mo(\br_2)} \right. \\
 \left. +
\frac{ \Delta n(\br_1)\Delta n(\br_2)}{4n\mo(\br_1)n\mo(\br_2)}- \frac{ (\Delta n(\br_1))^2}{8 (n^m(\br_1))^2}
- \frac{ (\Delta n(\br_2))^2}{8 (n^m(\br_2))^2}\right).
\end{multline}
The second order correction is of course small when $\Delta n$ is small, but it is also small when the ratio between the real and the model densities is similar in $\br_1$ and $\br_2$, which is a less severe requirement that may be satisfied by a non-homogeneous model.

\subsubsection{Connector approximations.}
In the connector philosophy, we do not want to expand the result around some point, but  optimize the zero-order, meaning that we find a model system that gives directly the correct result. As  explained above, one way to \textit{find} the parameters of this model is to use a low-order expansion. So the expansion is not used to produce directly the final result, but only to determine the parameters of the model, from which the result is taken. 
We will examine the connector approximation \textit{for the single-electron case}, where we do not need it of course, but where we can hope to learn something about connector approximations based on expansions.

The DM for the single homogeneous electron is peculiar, because its density matrix is linear in the density: it is simply the density itself. Therefore, the first-order expansion is exact in the homogeneous model, so $(O_{\rm approx}\mo)^{-1}=(O\mo)^{-1}$ and therefore  $Q^c=(O\mo)^{-1}O_{\rm approx}[Q]$, where $O\mo$ is $O$ on the subspace of model densities. The final result is then $O[Q^c]=O\mo[Q^c]=O_{\rm approx}[Q]$, which is equal to the first-order approximation itself: since the approximation equals the exact solution in the model, there is no error cancelling, and nothing is gained by the connector. 
In order for the first-order connector to be useful in the single-electron example, the model system must therefore be inhomogeneous.

In the following, we will test different approximations, including various connector approximations, for single-electron systems. Instead, in Subsec. \ref{subsec:many-electrons} we will use the \textit{exact} single-electron results as \textit{approximation} for the CT approach \textit{to the many-electron case}. 

 \subsubsection{Performance of the approximations in the single-electron case.}
 
 Although the single-electron case can be solved exactly, testing approximations to it will give guidelines for the many-electron case. We therefore start with a
numerical illustration for a single electron with density  $n(\br)=A_1 \cos(a_1 \br)+A_2 \cos(a_2 \br)+A_3 \cos(a_3 \br) + B $  where $\mathbf{a}=a \mathbf{\hat{x}}$ in a cube of side length  $2 \pi$.

\begin{figure}[!th]
\begin{subfigure}{\columnwidth}
  \includegraphics[width=0.5\linewidth]{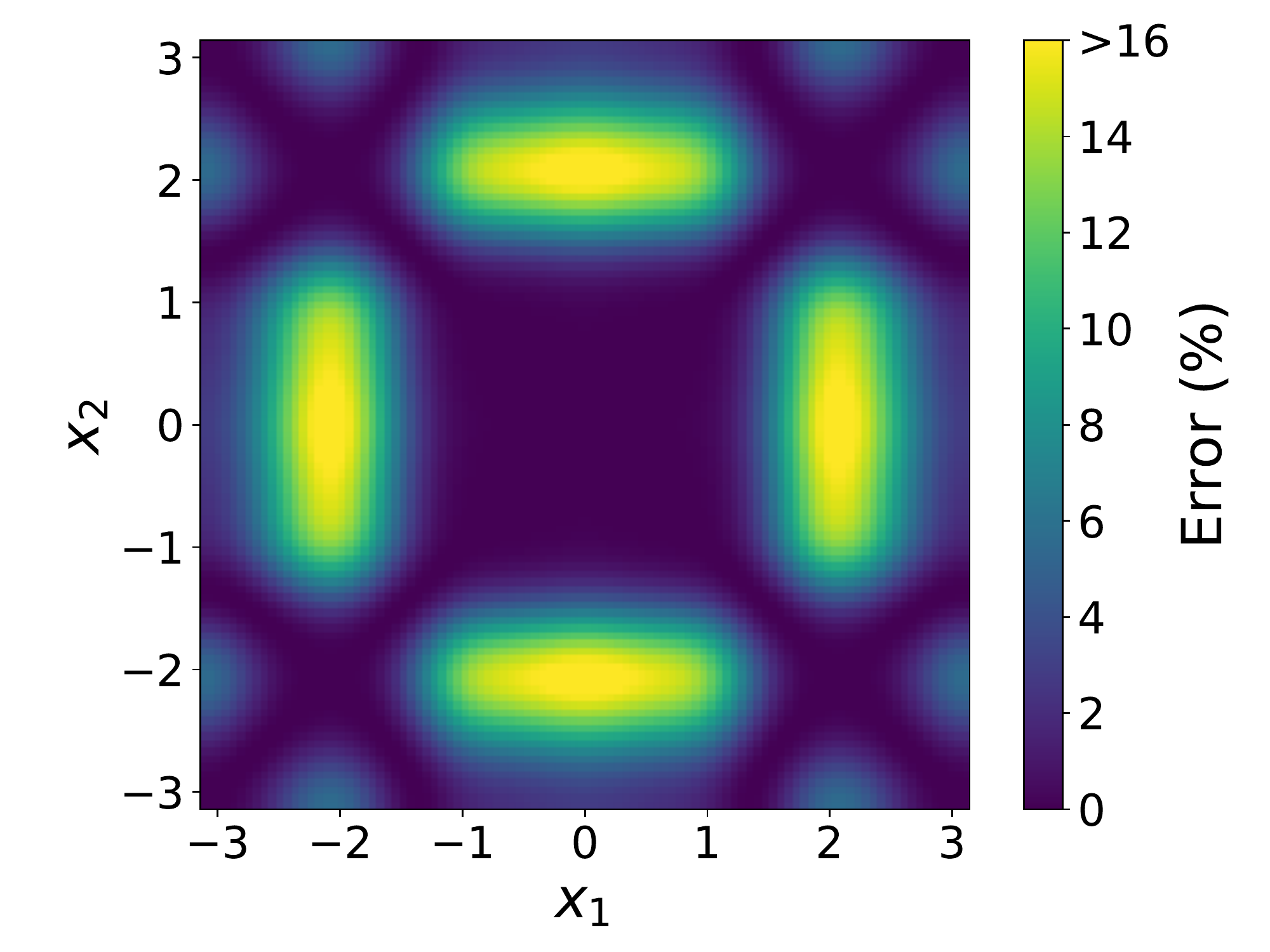}  
   \includegraphics[width=0.5\linewidth]{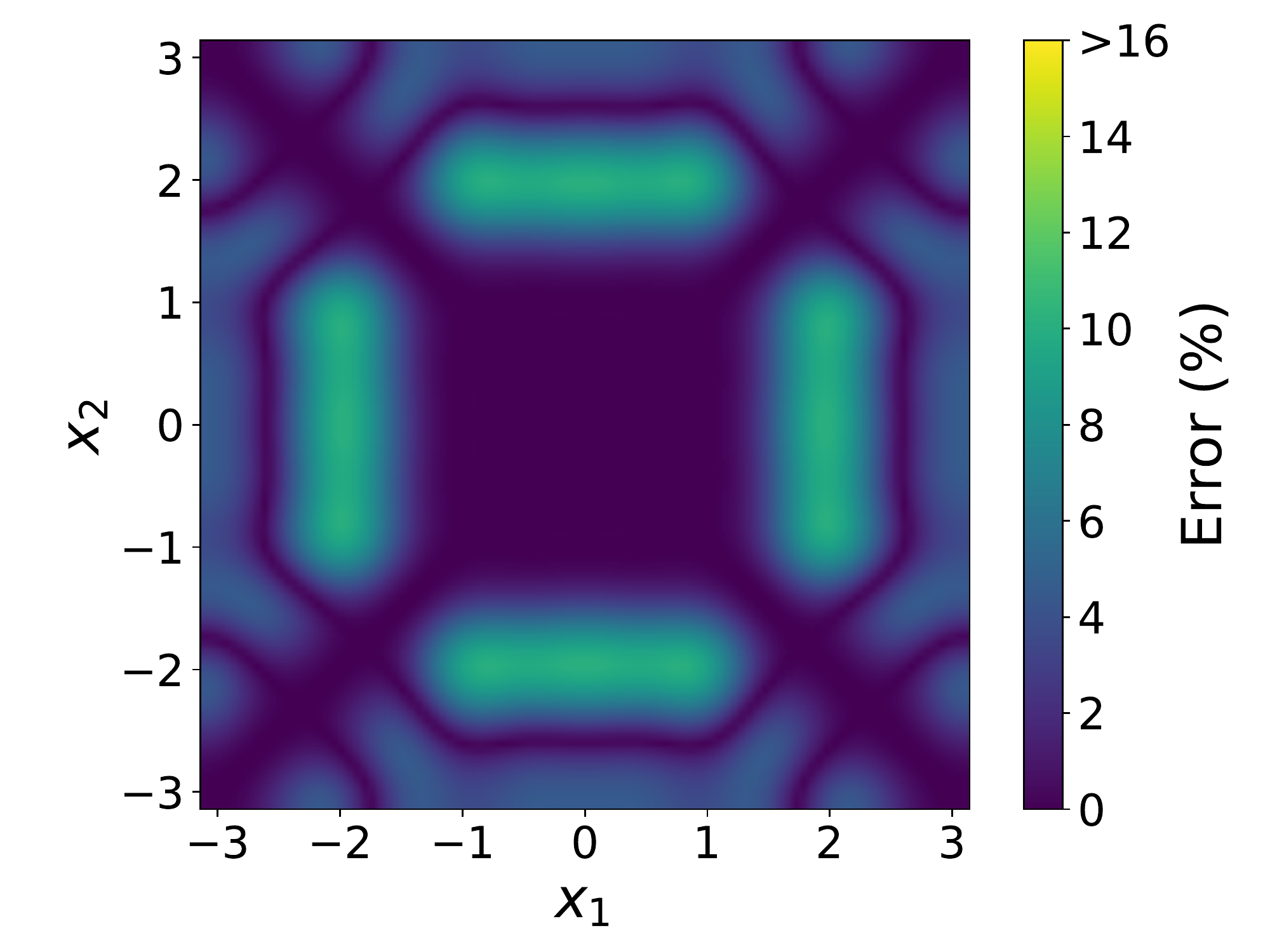} 
  \caption{First order approximation.(Left) Direct approximation. Mean relative error (MRE) = 4.33\% (Right) CT fixing $A^c=A_1$ and $a^c=a_1$. MRE =2.88\% .}
  \label{fig:expansions-first}
\end{subfigure}

\begin{subfigure}{\columnwidth}
  \centering
  \includegraphics[width=0.49\linewidth]{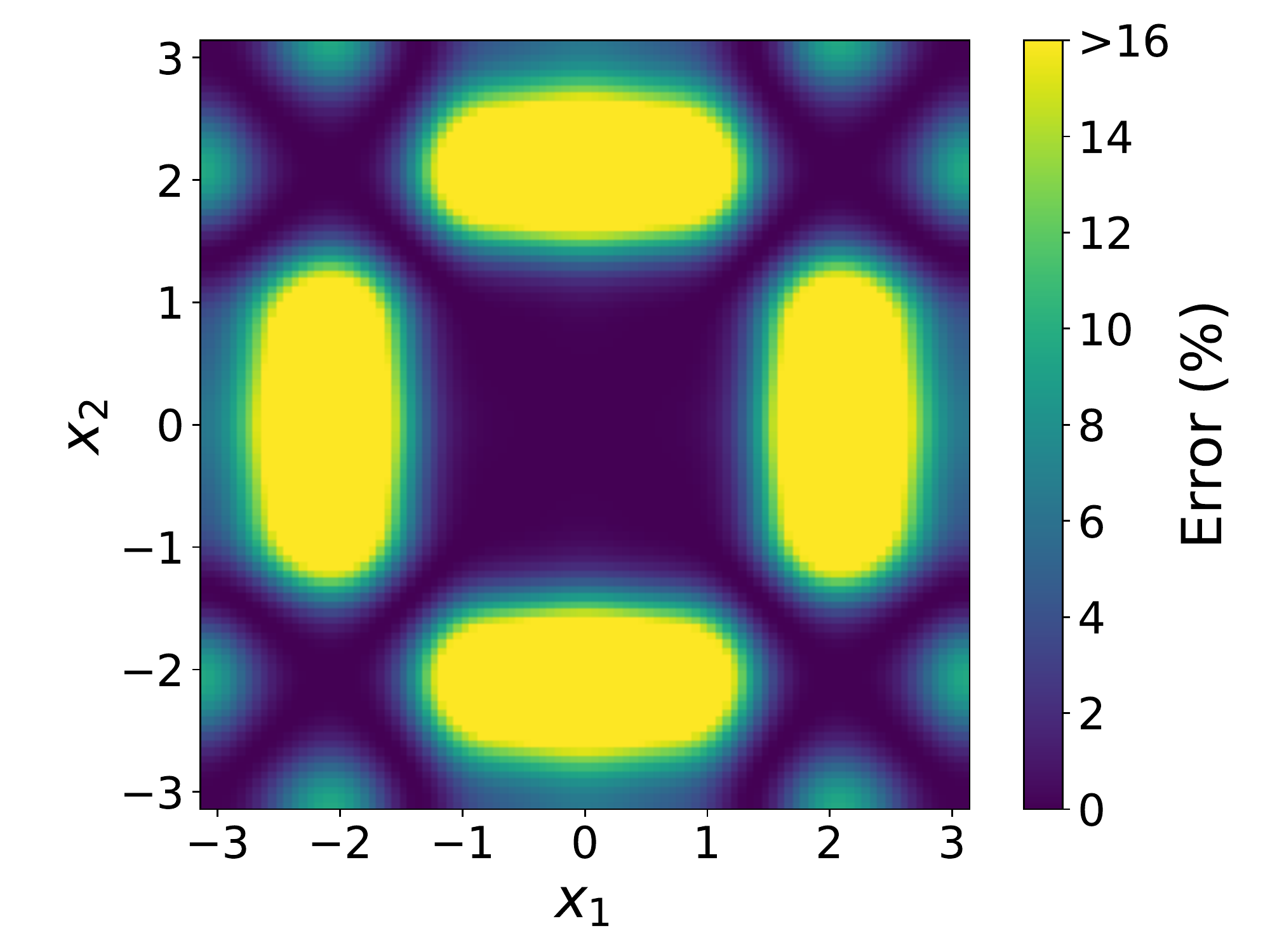}
  \includegraphics[width=0.49\linewidth]{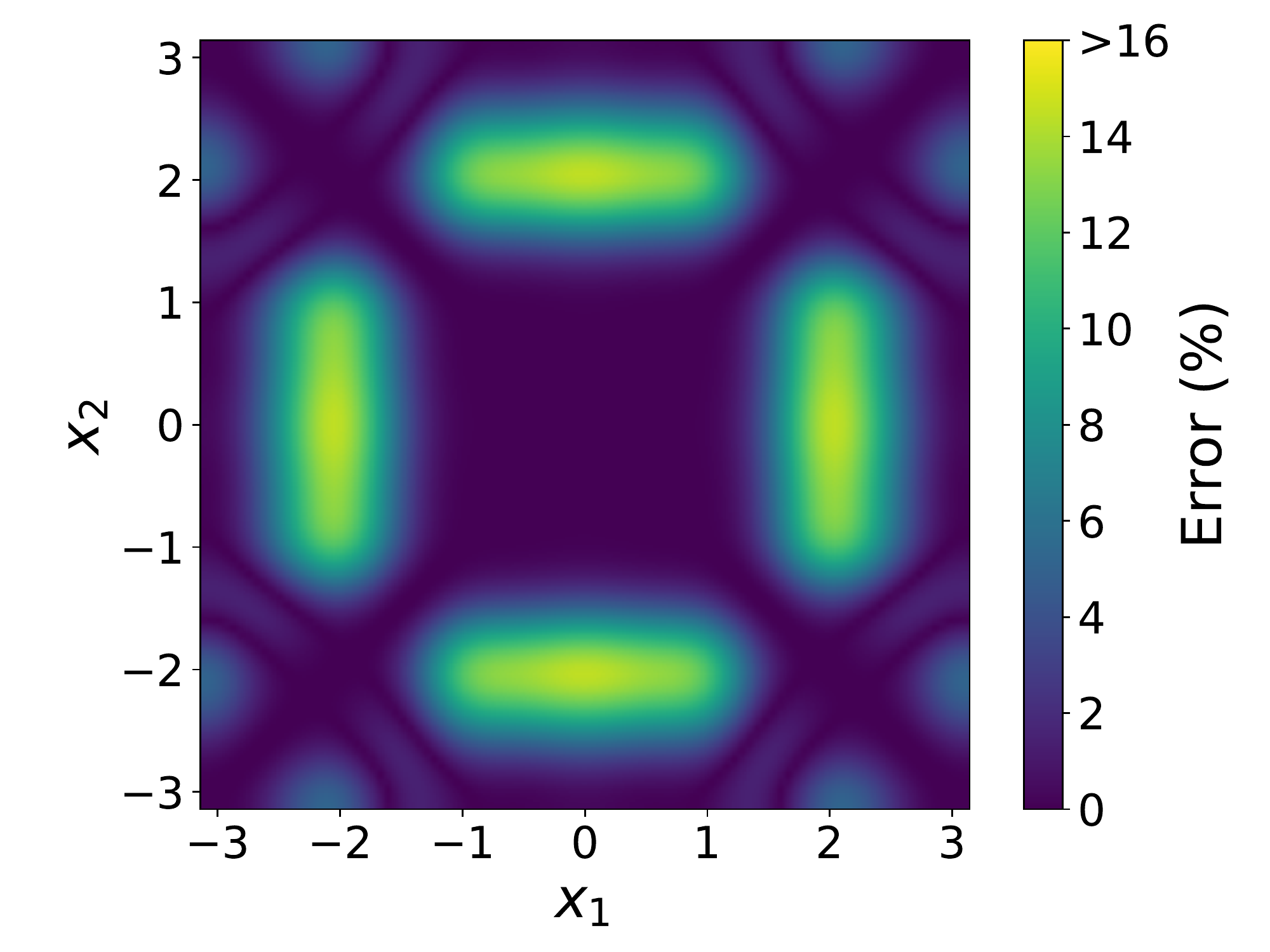}  
  \caption{(Left) Second order approximation starting from $n\mo=B$. MRE = 8.48\%
  (Right) First-order CT fixing $A^c=(A_1+A_2+A_3)/3$ and $a^c=a_1$. MRE = 3.51\%}
  \label{fig:expansions-second}
\end{subfigure}

\begin{subfigure}{\columnwidth}
  \centering
  \includegraphics[width=0.49\linewidth]{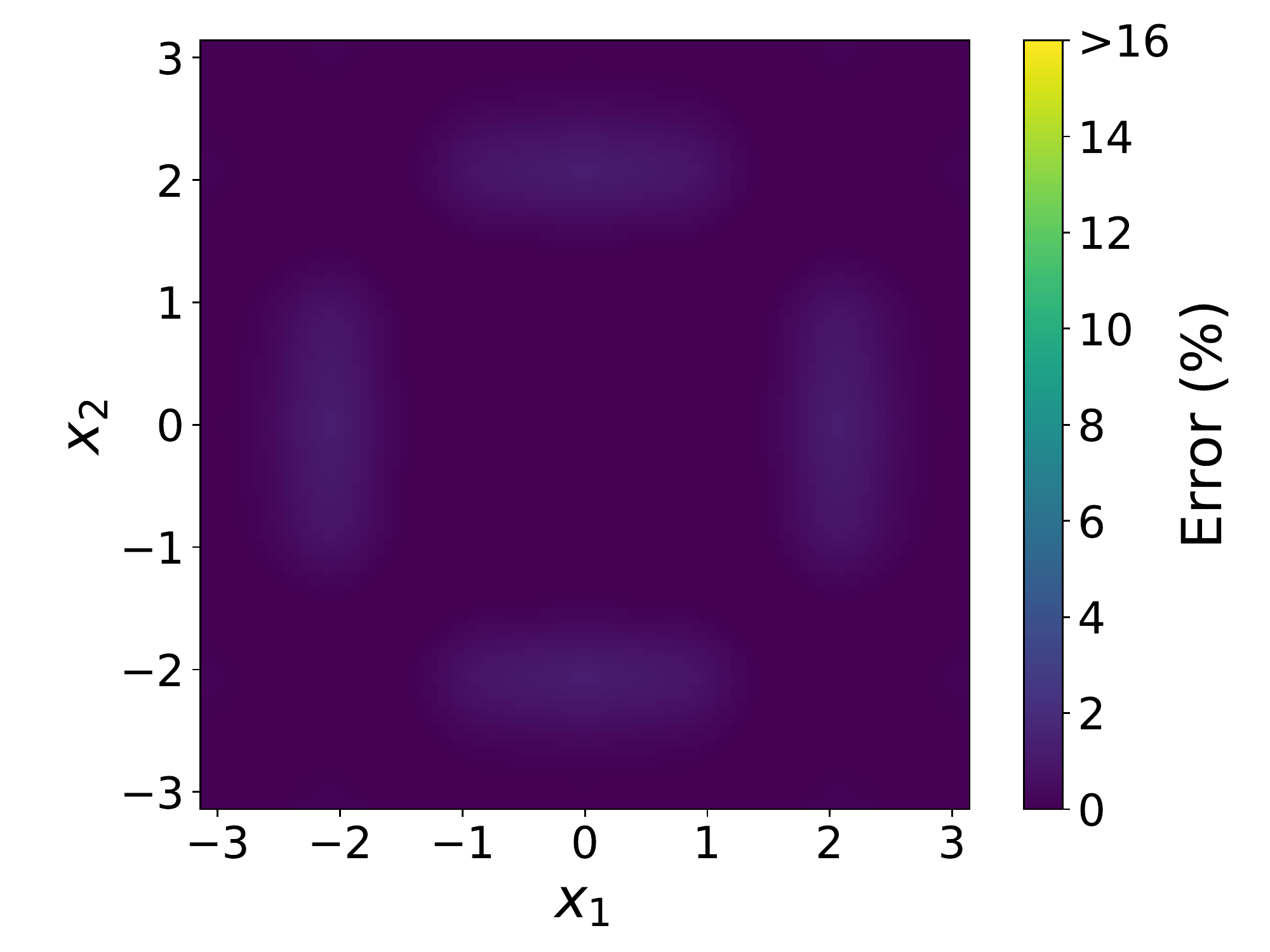}  
  \includegraphics[width=0.49\linewidth]{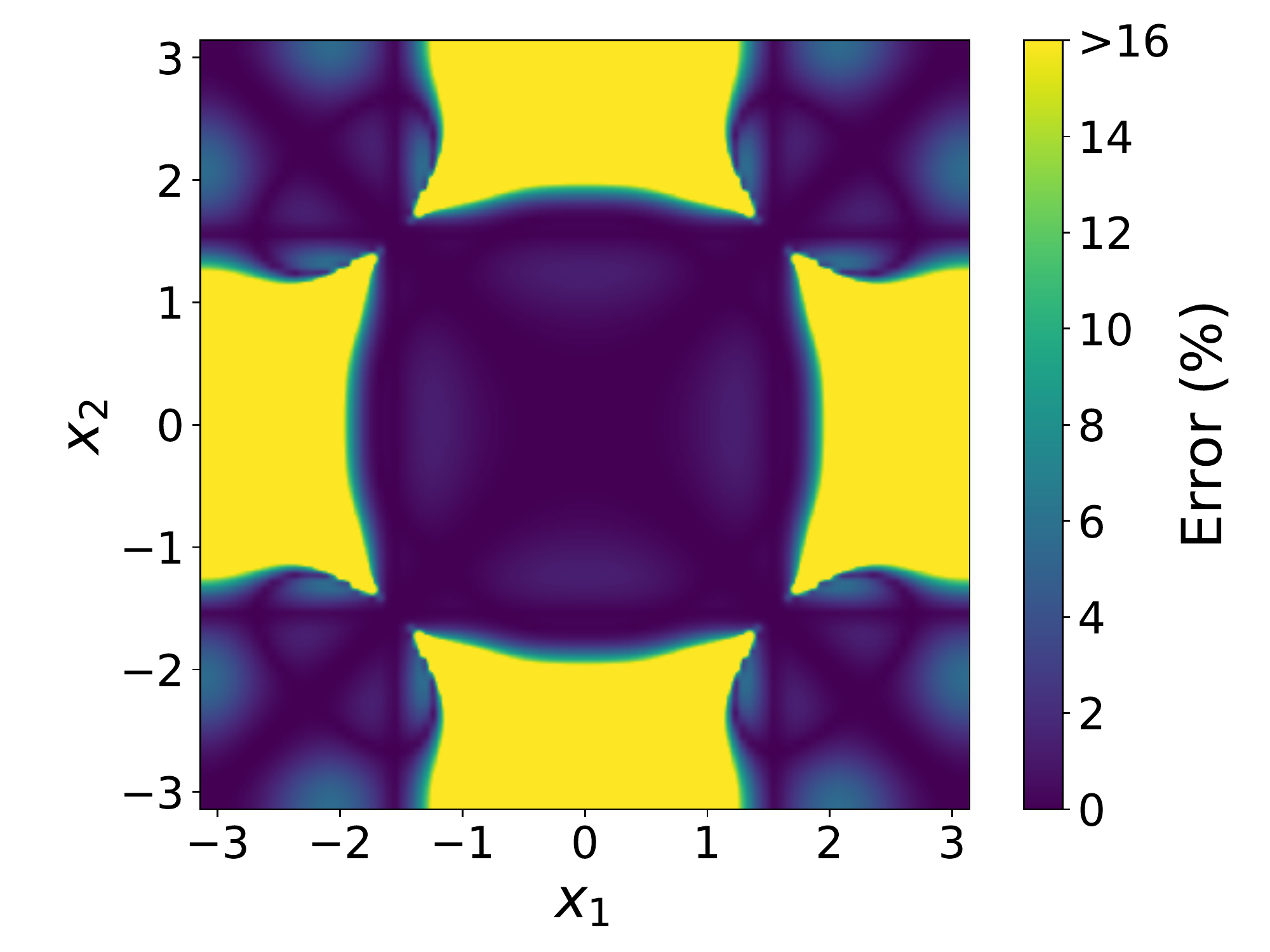} 
  \caption{(Left) As (b), with $n\mo=0.5(n(\br_1)+n(\br_2))$. MRE = 0.19\% \\
  (Right) First-order CT fixing $B^c=B$ and $a^c=a_1$. MRE = 27.95 \%}
  \label{fig:expansions-third}
\end{subfigure}

\caption{Relative error of approximations to the density matrix for a system with $n(\br)=A_1 \cos({\bf a}_1 \br)+A_2 \cos({\bf a}_2 \br)+A_3 \cos({\bf a}_3 \br) + B $,  where $\mathbf{a}=a \mathbf{\hat{x}}$, with $A_1=2$, $A_2=1$, $A_3=0.5$, $a_1=1$, $a_2=2$, $a_3=3$ and $B=3.1$ . Left column: first and second order expansions. Right column:  connector theory (CT) approximations based on first-order expansion around a homogeneous system, using for the connector a model with density $n\mo(\br)=A \cos(a\mathbf{\hat{x}} \br)+ B $. In principle $A$, $a$ and/or $B$ can be varied to connect the real and the model system; the results shown in the different panels are obtained with different choices.
}
\label{fig:expansions}
\end{figure}

To study the performance of different approximations, we show in Fig. \ref{fig:expansions} the relative error of the approximate density matrix with respect to the exact one, with $\br_1$ and $\br_2$ in direction $\hat {\bf x}$. 
  To characterize the results in a single number, the captions also show the mean relative errors (MRE). 
The left column displays the result of first- and second-order expansions  around different starting homogeneous densities. The same scale has been imposed to all results, which implies that in the bright yellow regions the error exceeds the maximum error set by the scale. The upper panel is the first order result. The error is largest in points $(\br_1,\br_2)$ for which the density is very small in one point and
and large in the other, which may lead to $\Delta n(\br_i)/n\mo\geq 1$. 
As pointed out above, the first-order result does not depend on the homogeneous starting point. The next two panels are second order results, starting from the mean density of the system, or from $n\mo_{\br_1\br_2}=0.5(n(\br_1)+n(\br_2))$, respectively. In the first case, the error is larger than that of the first order result, with maxima where one of the two densities is small, i.e. in regions where already the first order is problematic. In the second case instead, where a different homogeneous system is chosen for each pair $(\br_1,\br_2)$,
the result improves significantly.

The right column of Fig. \ref{fig:expansions} shows the result of the first order connector approximation. As explained above, we have to use an inhomogeneous model for a meaningful connector approximation based on a first-order expansion. We choose a model system with a density that is constrained to be much simpler than that of the real system: we allow only one Fourier component, $n\mo(\br)=A\mo \cos(a\mo \br)+ B\mo $, with parameters $A\mo$, $a\mo$ and $B\mo$ that can be varied to match the connector equality. To approximate this equality, as one would do in a real material, we use a first order expansion around $A\mo=0$ and $A_i=0$, i.e. a homogeneous density for the real and the model DM: this is the same approximation as the one that gave the first panel in the left column, but now, it is used within CT.

The three model parameters $A\mo$, $a\mo$ and $B\mo$ cannot be uniquely defined by the connector condition $O[Q]=O[Q\mo]$. This require to impose additional constraints and
solve for the remaining free model parameter(s), to finally obtain the set $A^c,a^c,B^c$. Since the connector approach has never been studied for such a case, it is interesting to explore what is the most promising way to set the model parameters. 

Fig. \ref{fig:expansions} shows that the CT result depends strongly on the way the parameters are set: the best result is obtained when the average density of the model, $B^c_{\br_1\br_2}$, is used as free parameter to optimize each pair of points $(\br_1,\br_2)$, while the amplitude and periodicity of the dominant Fourier component equals that of the real system. As can be seen by comparing the left and right panels of Fig. \ref{fig:expansions-first},  using the same first-order expansion the CT result is clearly superior with respect to the direct approximation while, when the model results are tabulated, the workload to calculate a real system is the same in both cases. The CT result worsens when the amplitude of the first fourier component is set to an average value, but it is still better than the direct approximation. The worst results are obtained when the average density is kept fixed, and the connector is set by the amplitude of the oscillation, as shown in the last panel. In all cases, the worst results are obtained when one of the densities is small. Note that in the last panel the CT equation would even yield regions with negative densities; in those cases, the density matrix has been set to zero. 
The best result is the second-order expansion in panel (c) in the left column; however, it would require in practice a higher computational effort. 
When the average density $B$ is much lower, e.g., $B=1$, results become worse. Still, for $n\mo=(n_{\rm max}+n_{\rm min})/2$ the expansion converges, though more slowly. 
 Altogether, the results indicate that one may obtain practicable approximations using expansions, even around a homogeneous system, and that in particular
 CT combined with a low-order expansion is a promising direction.

\subsection{The many-electron density matrix.}
\label{subsec:many-electrons}

Real systems cannot be solved analytically in general, but we can try to use insight from the single-electron case to approximate the density matrix of real systems\footnote{Also in the asymptotic region $\br_1,\br_2\rightarrow\infty$ of a finite system the density matrix $n(\br_1,\br_2)$ behaves like\cite{March1982} $\sqrt{n(\br_1)n(\br_2)}$ .}. We will do this in the following for the non-interacting, or Kohn-Sham, case.

\subsubsection{Nearsightedness of the many-electron density matrix.}
The perfect generalized nearsightedness of the density matrix is destroyed as soon as one has a system with more than one orbital,
even without interactions. To understand this better, we will look at the first-order expansion. In the case of one orbital, nearsightedness came through cancellations in ${^3\chi_0\mo}(\chi_0\mo)^{-1}$: the crucial point was  the cancellation of an occupied orbital in the integral of \eqref{eq:chi0-1-m} with \eqref{eq:3chi0-m}, which allowed one to integrate and have a cancellation of the energy denominators transition by transition, and finally use the completeness relation yielding $\delta$-functions.  In the case of more orbitals, only when the different orbitals are localized in completely different regions of space and can therefore be treated separately, things will work in a similar way, and generalized nearsightedness will hold. Indeed, in that situation, in each region the density matrix is that of one electron, and off-diagonal elements connecting different regions vanish. Note again that general nearsightedness of the density matrix does not require $\chi_0\mo (\br_1,\br_2)$ to be ultra short ranged in $|\br_1-\br_2|$, as discussed earlier.

\subsubsection{The many-electron density matrix of real systems: CT approximations}

One might expect that generalized nearsightedness also holds in the many-electron system, as a first guess. To put things on a firmer ground, CT approximations might be helpful to improve the results when the nearsightedness breaks down, since they are able to interpolate between the nearsighted and farsighted cases. We will concentrate on CT in the following.

First, note that expanding is only one possibility to realise the connector strategy, which requires more generally to \textit{do the same approximation on the real and the model system}. In order to illustrate the CT for real systems in the probably simplest possible case, and in order to draw maximum benefit from the analytic single-electron case, we will in the following use the \textit{single electron as approximation to build the connector}.  

We will study two real systems, bulk silicon\cite{Wyckoff1963} and solid $^{3}$He\cite{Schuch1961}, on the Kohn-Sham level. Our model system will be homogeneous, characterized by its density. In the single electron approximation, the connector equality reads 
\begin{eqnarray}
n^c_{\br_1\br_2} = \sqrt{n(\br_1)n(\br_2)}.
\end{eqnarray}
The density matrix $n(\br_1,\br_2)$ in a pair of points is then obtained in the CT approximation by evaluating the density matrix of the HEG with homogeneous density $n^c_{\br_1\br_2}$.  Using the HEG with geometric mean density is quite intuitive, and has been used e.g. as ingredient for an energy functional describing van der Waals interaction \cite{Rapcewicz1991}.

\begin{figure}[!th]
\begin{subfigure}{\columnwidth}
  \includegraphics[width=0.5\linewidth]{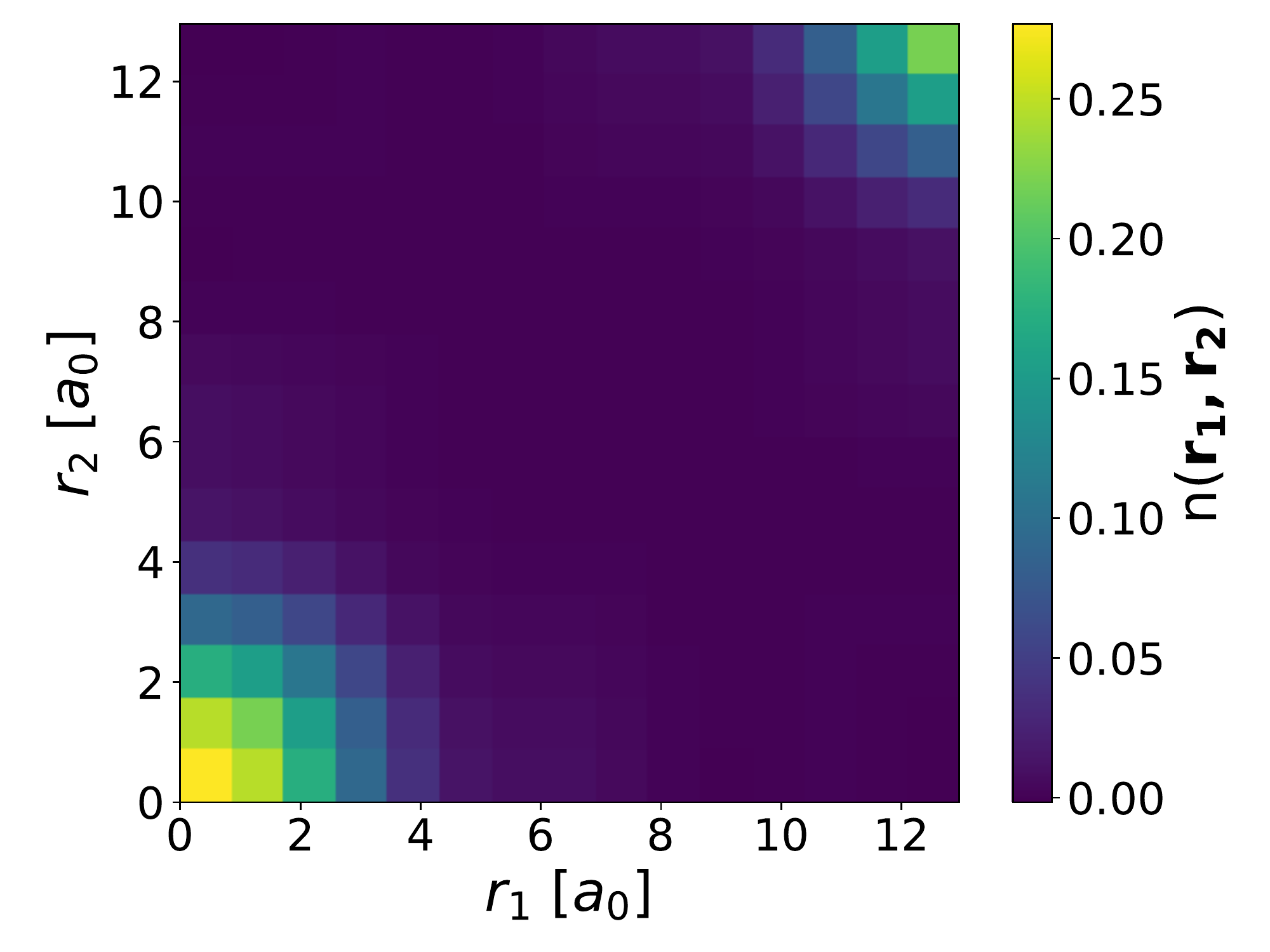}  
   \includegraphics[width=0.5\linewidth]{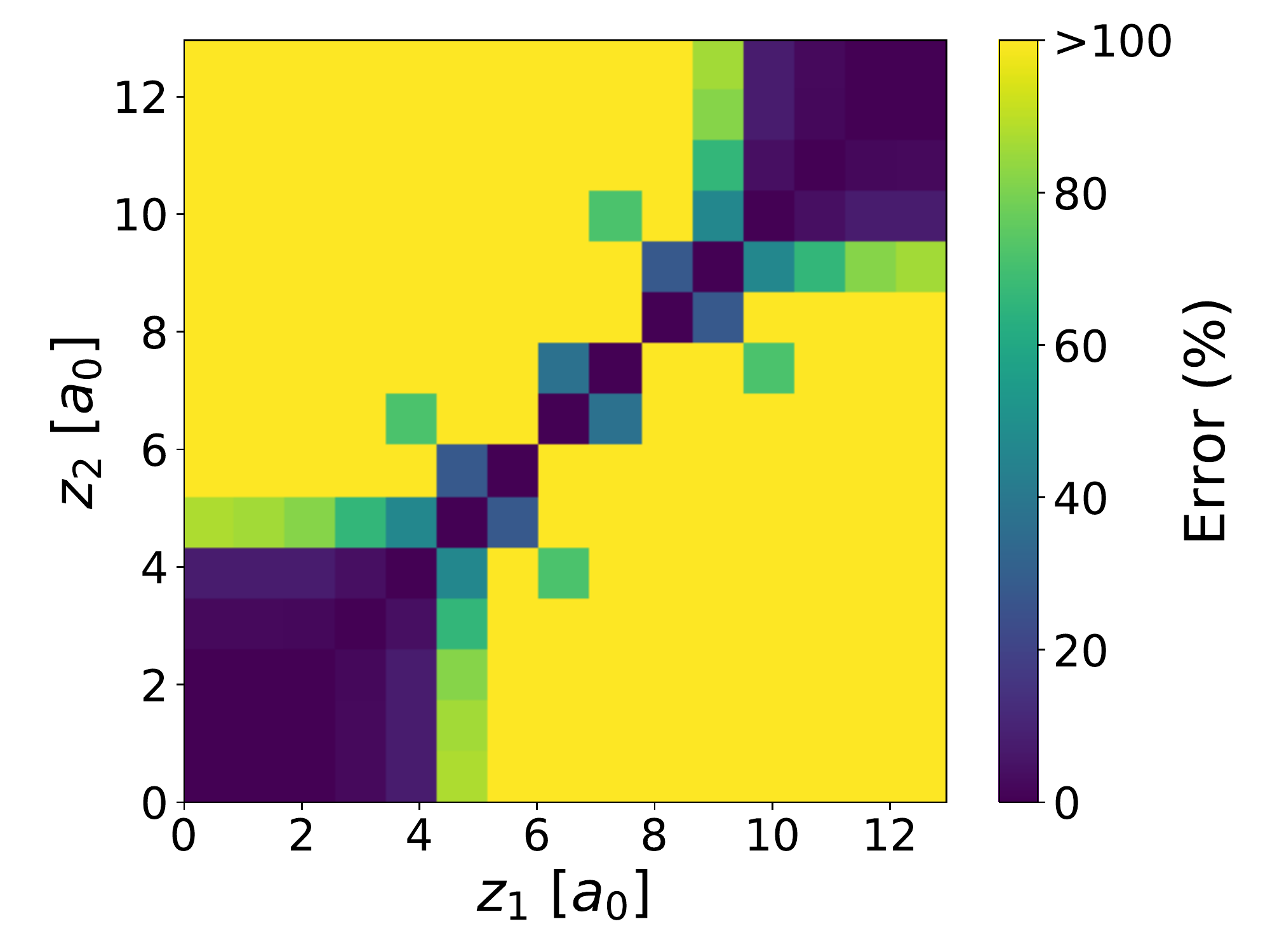} 
  \caption{(Left)  Kohn-Sham DM.    (Right) Direct single-electron approximation. }
  \label{fig:He_direct}
\end{subfigure}

\begin{subfigure}{\columnwidth}
  \centering
  \includegraphics[width=0.49\linewidth]{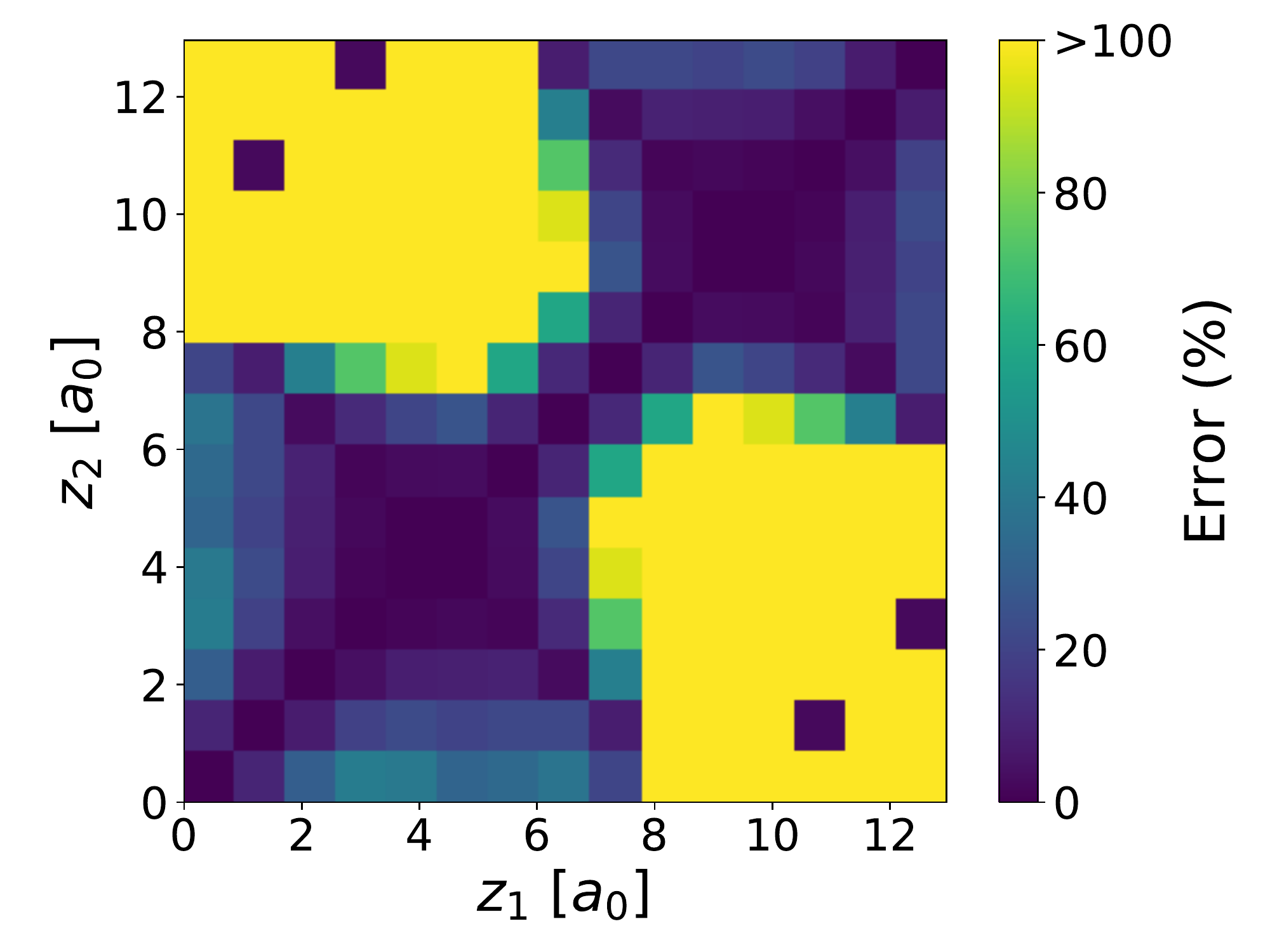}
  \includegraphics[width=0.49\linewidth]{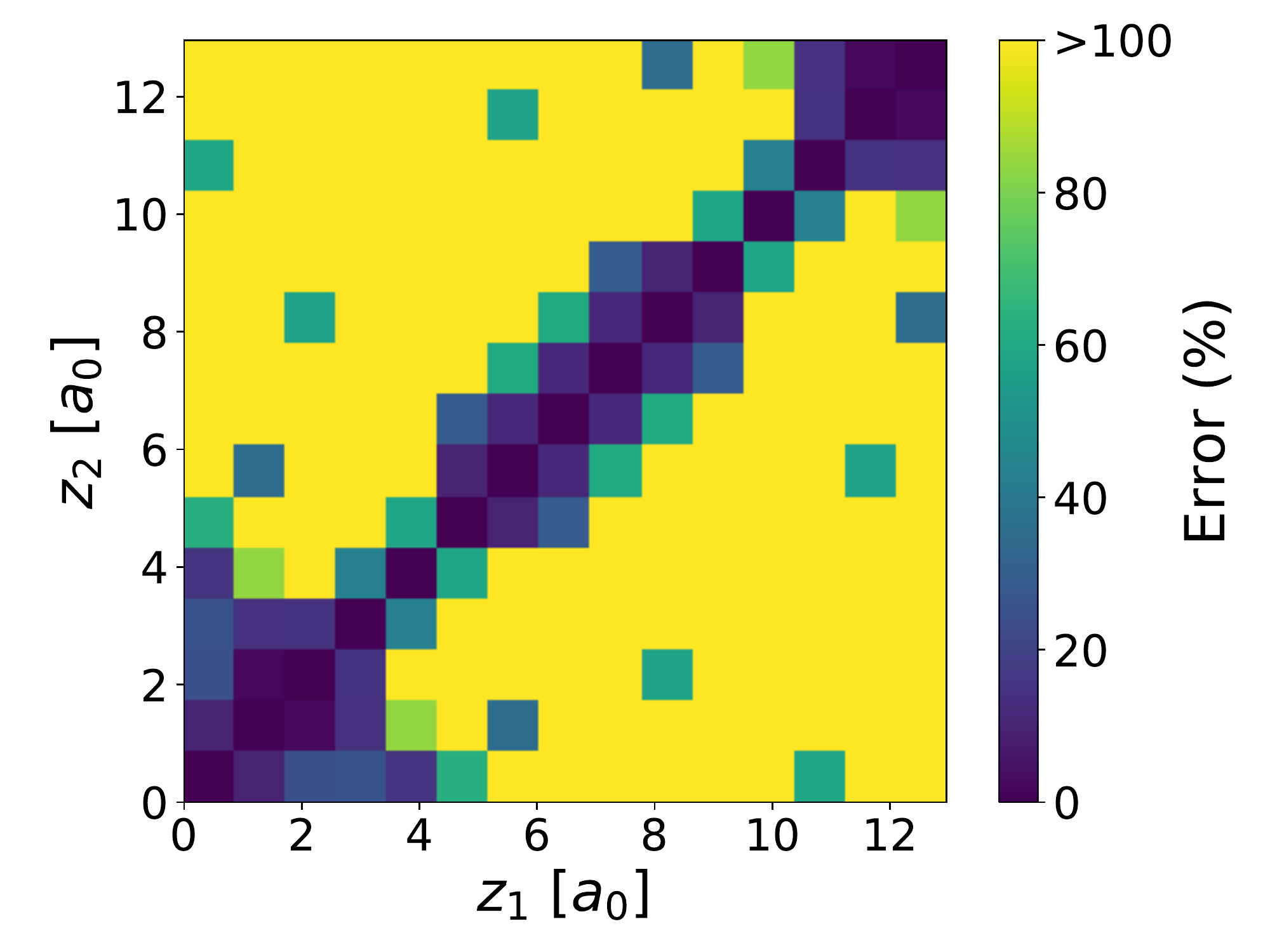}  
  \caption{(Left)CT  with single-electron approximation\\
  (Right) CT with $n\mo_{\br_1\br_2}=0.5(n(\br_1)+n(\br_2))$}
  \label{fig:He_ct}
\end{subfigure}

\caption{DM and relative error of approximations to the DM for solid helium along the [1,1,1] direction.}
\label{fig:He}
\end{figure}

\begin{figure}[!th]
\begin{subfigure}{\columnwidth}
  \includegraphics[width=0.5\linewidth]{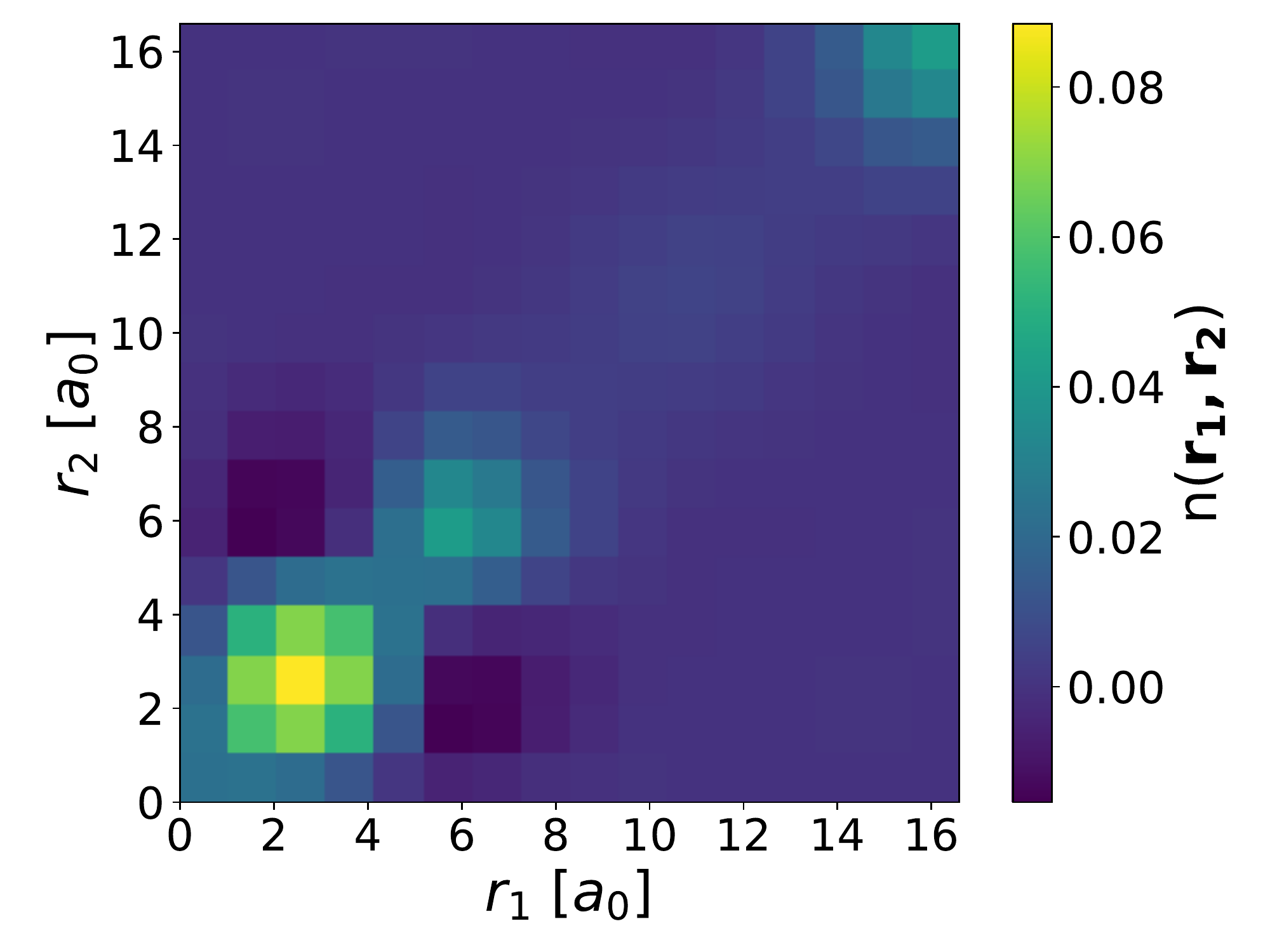}  
   \includegraphics[width=0.5\linewidth]{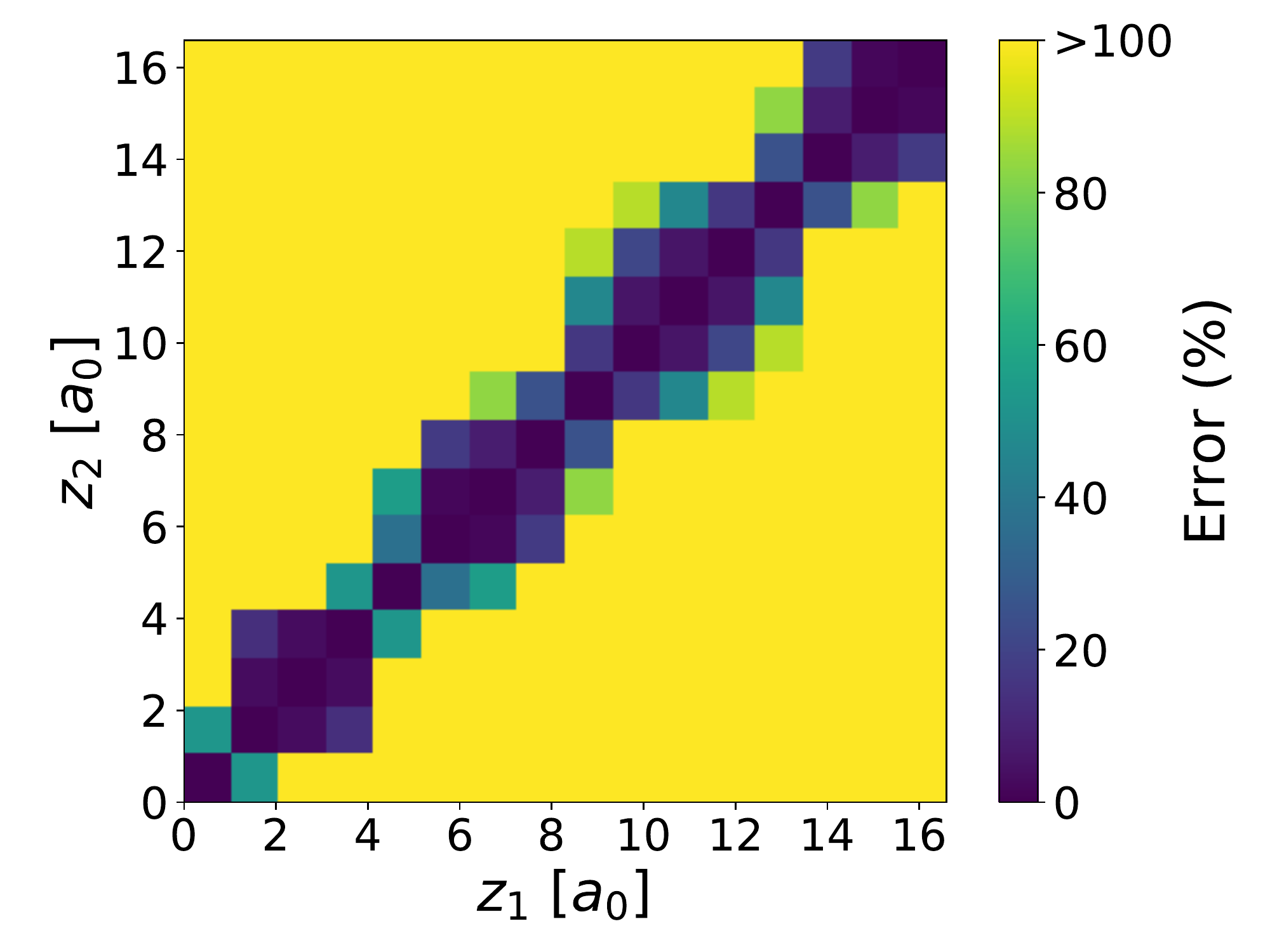} 
  \caption{(Left)  Kohn-Sham DM.    (Right) Direct single-electron approximation}
  \label{fig:Si_direct}
\end{subfigure}

\begin{subfigure}{\columnwidth}
  \centering
  \includegraphics[width=0.49\linewidth]{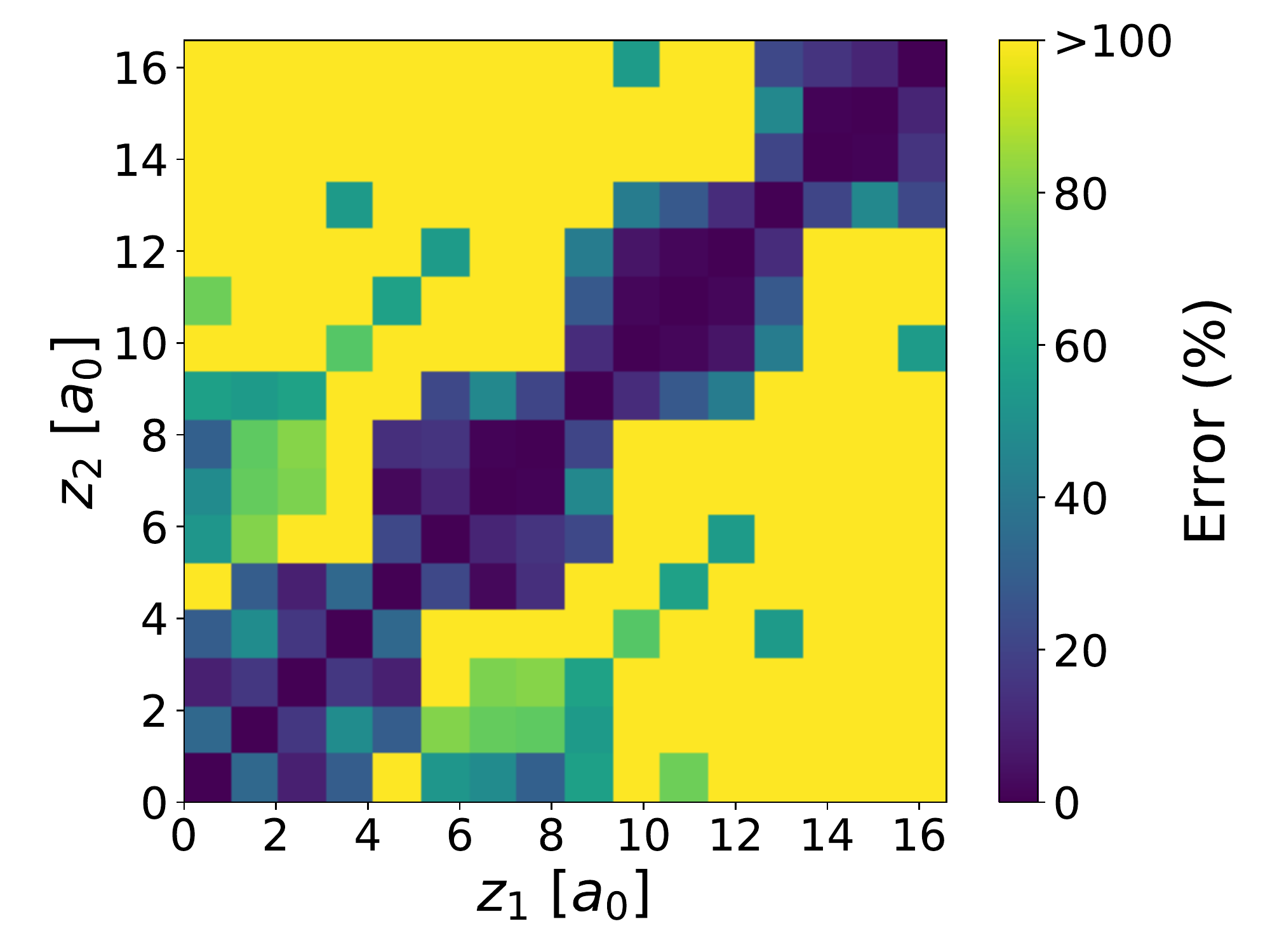}
  \includegraphics[width=0.49\linewidth]{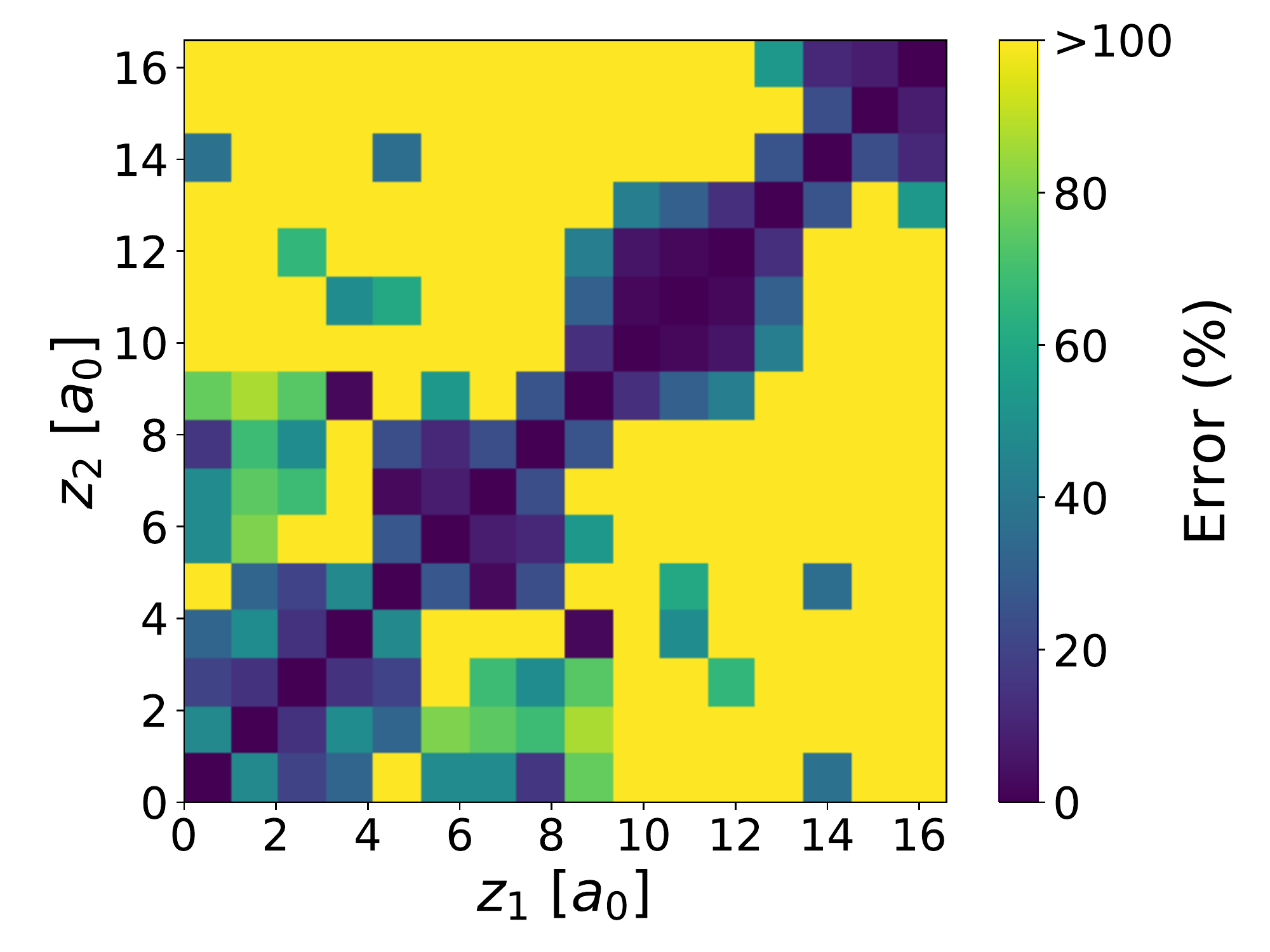}  
  \caption{(Left) CT  with single-electron approximation \\
  (Right) CT with $n\mo_{\br_1\br_2}=0.5(n(\br_1)+n(\br_2))$}
  \label{fig:Si_ct}
\end{subfigure}

\caption{DM and relative error of approximations to the DM for bulk silicon along the [1,1,1] direction.}
\label{fig:Si}
\end{figure}

Figs. \ref{fig:He} and \ref{fig:Si} show, for helium and silicon, respectively,  the Kohn-Sham density matrix $n(\br_1,\br_2)$, the direct single-electron approximation $n(\br_1,\br_2)\approx \sqrt{n(\br_1)n(\br_2)}$, and the CT approximation $n(\br_1,\br_2)\approx n(|\br_1-\br_2|;[n^c_{\br_1\br_2}])$. Figs. \ref{fig:He} and \ref{fig:Si} also show results of the CT approximation with  guesses for the connector extrapolated from the single electron.
The relative error on the exchange energy calculated over one unit cell is given in in Table \ref{tab:ex_err}.

\begin{table}[!ht]
 \centering
 \small
     \caption{Total error on exchange energy (in \%)} 
    \label{tab:ex_err}
    \begin{tabular*}{0.48\textwidth}{@{\extracolsep{\fill}}lrr}
    \hline  
     & He & Si \\
     \hline 
   Direct single-electron  & 167 & 575 \\
  Direct $0.5(n(\br_1)+n(\br_2))$  &1117  & 663\\
  Direct average density  & 1455 & 663 \\
  CT single-electron  & 51 & 6 \\
  CT $0.5(n(\br_1)+n(\br_2))$   & 134 & 6 \\
  CT average density & 833 & 10 \\
 \hline
    \end{tabular*}
\end{table}

Errors are now much larger than in the previous subsection, since the single-electron approximation is far from an infinite many-electron system, but trends are as before: CT significantly improves the results with respect to a direct use of the approximation. Again, errors in the exchange energy are smaller than the error in the density matrix itself. For silicon, which is closer to the HEG, i.e., closer to the model used by CT, even with this very simple approximation the error might even be acceptable for some applications. Also for helium the connector improves the results, but compared to silicon the direct approximation is better and the CT result is worse, because helium is closer to the single electron than to the HEG.

\subsection{Comments on the interacting density matrix}
Finding simple density functionals for the non-interacting density matrix is an important topic, since it could for example allow one to perform orbital-dependent Kohn-Sham calculations, or speed up calculations based on hybrid functionals. 
The practical part of the present work was focused on that topic, and to go beyond would exceed its frame. Therefore, we simply complement the discussion by a few remarks concerning the interacting case.

As discussed in Sec. \ref{sec:balloon}, one could expand the interacting density matrix with respect to the interacting or the non-interacting density. While with the second choice one remains, besides the derivatives of the density matrix itself, with non-interacting response functions, the former choice has the advantage of yielding the exact diagonal, the density, to all orders. 

One may expect that the generalized nearsightedness is strongly weakened by interactions, since the full response $\chi$ feels the long-range Coulomb interaction, contrary to $\chi_0$. However, as explained earlier, nearsightedness is not based on the response function itself, but on the chain of response functions and their inverse, which gives some hope also in the interacting case. Moreover, using the CT one does not have to guess nearsightedness, but can let the connector set the pertinent range.   It seems therefore worthwhile to explore this route further in the future.

\section{Time non-local observables: spectral functions}
With the density matrix half way between the density and the one-body \GF ,
it is a unifying topic of DFT and GFFT. As we have seen, building density functionals for the DM seems to be within reach. It is, instead, much more difficult for the full \GF , and in the following we will briefly outline why. To understand the difficulty, it is again sufficient to look at the case of a single electron. Its \GF\ reads
\begin{equation}
    G_0(\br,\brp;\omega) = \frac{\phi_v(\br)\phi_v(\brp)}{\omega - \varepsilon_v -i\eta} + \sum_c \frac{\phi_c(\br)\phi_c(\brp)}{\omega - \varepsilon_c +i\eta}.
\end{equation}
This expression is not separable (and therefore, not nearsighted in the generalized sense) because of the sum over empty states. However, it is nearsighted in the generalized sense if one is only interested in electron removal spectra. In that case 
\begin{equation}
    G_0^{\rm hole}(\br,\brp;\omega) \equiv \frac{\phi_v(\br)\phi_v(\brp)}{\omega - \varepsilon_v -i\eta} = \sqrt{G_0^{\rm hole}(\br,\br,\omega)}\sqrt{G_0^{\rm hole}(\brp,\brp,\omega)}:
\end{equation}
this suggests to express the non-local \GF\ as functional of the local \GF . This very simple starting point may indicate a direction towards approaches like DMFT, where functionals of the local \GF\ are approximated using nearsightedness. Still, this is not an explicit density functional. If one examines the balloon expansion Sec. \ref{sec:balloon}, the difficulty becomes clearer. With respect to the DM, the only change is in the outmost response functions, which carry now two different times at their endpoints. 

In the one-electron case, and to first order in the density, let us look at the case  $t_1>t_2$ which corresponds to electron addition. The calculations proceed as before, but an additional phase factor $e^{-i\varepsilon_c(t_1-t_2)}$ appears in \eqref{eq:3chi0-m}.  This prevents us from using the completeness relation; after a frequency fourier transform the non-nearsightedness of the resulting contributions is
given by $\frac{\phi_c(\br_1)\phi_c(\br_6)}{\omega-\varepsilon_c-i\eta}\Delta n(\br_6)$, similarly for $\br_1\to \br_2$. The correction with respect to $G_0\mo$ obtained in this way is a non-nearsighted renormalization of the weight of each pole. This expression is near-sighted only for frequencies very far from all electron addition energies $\varepsilon_c$. In time space, this corresponds to a neglect of the phase factor, which is legitimate only for very short time differences, shorter than the inverse of the maximum addition energy. 
To see how the near-sighted result of the density matrix emerges, one can take the imaginary part of the correction in frequency space, and integrate over frequency.

Moreover, ${^3\chi_0\mo}$ has now an additional contribution, due to the fact that one can have $t_1>t_3>t_2$, which was not possible for $t_1=t_2$. In this case, ${^3\chi_0\mo}$ consists of only conduction states $c$ and 
one has to integrate $\frac{|\phi_c(\br_3)|^2\phi_{c'}(\br_3)}{\phi_v(\br_3)}$, where states $c'$ are conduction states that appear in $(\chi_0\mo)^{-1}$. Contrary to before, now the integral does not lead to $\delta_{cc'}$, and there is no indication for nearsightedness.
Another important point is that this contribution carries a factor $(t_1-t_2)e^{-i\varepsilon_c(t_1-t_2)}$, which leads to a double pole, $\frac{1}{(\omega-\varepsilon_c-i\eta)^2}$. 
Higher order corrections add higher order poles. This is a remnance of the usual expansion of a Dyson equation, where the kernel shifts the poles through a series of multiple poles. The convergence of such a series close to the poles is extremely slow, which indicates that a straightforward expansion of the \GF\  in the density is not convenient to get spectral functions.

Therefore, we can conclude that calculating spectral functions, or more generally, parts of \GF s that exhibit time differences, most likely requires functionals that have a significantly non-local dependence on the density.  Moreover, the straightforward balloon expansion in the density difference will have difficulties to shift the poles of the \GF\ from the model ones to the poles of the \GF\ of the real system. 
Indeed, usually the problem of shifting poles is overcome  when one expands the \textit{inverse} \GF , which is equivalent in the present case to expanding the external potential and, in the interacting case, the difference between the self-energies of the real and the model system, with respect to the density.

\section{Conclusions and Outlook}

 Building functionals of \GF s or of the density instead of the full many-body wavefunction is an efficient way to calculate observables, if appropriate functionals can be found. Here we have examined two ingredients that have been used succesfully, namely, expansions and the use of model systems. Functionals of the one-body \GF\ are most often known as expansions in the Coulomb interaction, and density functionals are often based on the use of the HEG as model system. These choices are, however, not a fatality. In particular, with today's computer and storage capacities it is interesting to explore whether the massive use of interacting model systems beyond the HEG could be a promising way to go, since this allows one in principle to calculate most of the interaction effects once and forever, and  the calculations for a variety of real materials would consist essentially in putting together the Lego pieces obtained from the model. The resulting functionals can be functionals of the model density and density differences, or functionals of the model \GF\ and its difference to the \GF\ of the real system. 

This direction is particularly appealing to build density functionals, since relatively simple expressions can be obtained. In order to examine what could be promising, we have concentrated on the density matrix as object to be approximated.
From our analysis and numerical results, we conclude that there is indeed hope to design practical approximations for the density matrix as explicit functional of the density. On one hand, expansions may converge fast for systems where the variations in the density are clearly smaller than the average density. The choice of the starting point of such an expansion is crucial. In the simplest case one expands around the HEG, and the best results are obtained when a different HEG is taken for each pair of points $(\br_1,\br_2)$ in $n(\br_1,\br_2)$, namely, a HEG with density $n\mo_{\br_1\br_2}=(n(\br_1)+n(\br_2))/2$. With respect to the expansions, the use of connector theory improves the results while requiring the same computational effort, and appears the way to go. It also allows one to use different approximations, not based on expansions, which might be helpful for systems with large variations with respect to the average density. 

Finally, neither expansions nor CT approximations are bound to the HEG as model system.  As an outlook, we have examined single elements of the density matrix of helium within connector theory and the single electron approximation. In Sec. \ref{sec:density-matrix} the CT calculations were done using the HEG as model system, and we can directly compare the results to what is obtained when the model is instead a system with periodic density, described by one Fourier component in each direction. Such a model is described by a limited number of parameters, and could still be tabulated if it turns out to be useful. Using the single electron approximation and imposing the CT equality requires $n(\br_1)n(\br_2)=n\mo(\br_1)n\mo(\br_2)$, which can be fulfilled in many ways. The most intuitive is the generalized LDA, $n(\br_i)=n\mo(\br_i)$ for $i=1,2$. We have examined a few elements of the density matrix where the real and model density in the two points are similar, and found that the inhomogeneous model improves the CT results by an order of magnitude. This is to be expected, since CT heavily relies on the model and becomes exact in the limit where the model equals the real system. However, it is encouraging to see that the improvement happens so fast.

For our discussions, we have chosen crude approximations, namely, low-order expansions or the single electron approximation. This has allowed us to highlight differences. One could do much better without losing efficiency; for example, for a periodic system the single electron approximation could be replaced by that of a periodic array of single electrons. We expect that in this way the errors on quantities such as the exchange energy could be brought into an acceptable range quite easily, and that large-scale calculations could benefit from tabulated model results, in the same way as DFT in its beginnings has enormously profited from the QMC results in the HEG, through the LDA.

\section*{Conflicts of interest}
There are no conflicts to declare.


\balance


\bibliography{biblio} 
\bibliographystyle{rsc} 

\end{document}